\documentclass{aa}
\usepackage[varg]{txfonts}
\usepackage{graphicx}
\usepackage{ulem}
\usepackage{color}
\usepackage{ascmac}

\usepackage{natbib}
\usepackage{bm}
\bibpunct{(}{)}{;}{a}{}{,} 

\definecolor{orange}{rgb}{ 0.95, 0.60, 0}

\begin{document}

\title{Effects of global gas flows on type I migration}
\subtitle{}
\author{Masahiro Ogihara\inst{\ref{inst1}}
\and Eiichiro Kokubo\inst{\ref{inst1}}
\and Takeru K. Suzuki\inst{\ref{inst2}}
\and Alessandro Morbidelli\inst{\ref{inst3}}
\and Aur\'{e}lien Crida\inst{\ref{inst3},\ref{inst4}}
}
\institute{Division of Theoretical Astronomy, National Astronomical Observatory of Japan, 2-21-1, Osawa, Mitaka, 181-8588 Tokyo, Japan \email{masahiro.ogihara@nao.ac.jp}\label{inst1}
\and School of Arts \& Sciences, University of Tokyo, 3-8-1, Komaba, Meguro, 153-8902 Tokyo, Japan\label{inst2}
\and Laboratoire Lagrange, Universit\'e C\^ote d'Azur, Observatoire de la C\^ote d'Azur, CNRS,
Blvd de l'Observatoire, CS 34229, 06304 Nice Cedex 4, France\label{inst3}
\and Institut Universitaire de France, 103 Boulevard Saint-Michel, 75005 Paris, France\label{inst4}
}


\abstract 
{
Magnetically-driven disk winds would alter the surface density slope of gas in the inner region of a protoplanetary disk $(r \lesssim 1 {\rm au})$. This in turn affects planet formation. Recently, the effect of disk wind torque has been considered with the suggestion that it would carve out the surface density of the disk from inside and would induce global gas flows (wind-driven accretion). 
} 
{We aim to investigate effects of global gas flows on type I migration and also examine planet formation.} 
{A simplified approach was taken to address this issue, and \textit{N}-body simulations with isolation-mass planets were also performed.} 
{In previous studies, the effect of gas flow induced by turbulence-driven accretion has been taken into account for its desaturation effect of the corotation torque. If more rapid gas flows (e.g., wind-driven accretion) are considered, the desaturation effect can be modified. In MRI-inactive disks, in which the wind-driven accretion dominates the disk evolution, the gas flow at the midplane plays an important role. If this flow is fast, the corotation torque is efficiently desaturated. Then, the fact that the surface density slope can be positive in the inner region due to the wind torque can generate an outward migration region extended to super-Earth mass planets. In this case, we observe that no planets fall onto the central star in \textit{N}-body simulations with migration forces imposed to reproduce such migration pattern. We also see that super-Earth mass planets can undergo outward migration.} 
{Relatively rapid gas flows affects type I migration and thus the formation of close-in planets.}
\maketitle

\section{Introduction}
According to numerous recent studies, global disk structures and evolution are more complicated than previously considered \citep[e.g.,][]{fromang_etal11,flock_etal11,flock_etal12,dzyukevich_etal13,parkin_biknell13,suzuki_inutsuka14,bethune_etal17}. Recently, magnetically-driven disk winds have been shown to have a crucial role in the disk evolution. 

\citet{suzuki_inutsuka09} performed local three-dimensional magnetohydrodynamic (MHD) simulations and showed the existence of turbulence-driven disk winds, in which vertical gas outflows are triggered by magnetorotational instability (MRI) turbulence. \citet{suzuki_etal10} further performed MHD simulations and found that turbulence-driven disk winds are observed even in MRI-inefficient disks, because disk winds can be driven near the surface region where magnetic fields are more turbulent. They also calculated the global disk evolution with mass loss (vertical gas outflows) due to disk winds, and found that mass loss due to disk winds alter the surface density distribution of gas disks, which can affect planet formation process in several ways. The dynamics and evolution of dust grains can be affected; small grains that are well coupled to the gas component are blown upward and away by the gas drag force, while larger grains that are loosely coupled to the gas remain in the disk \citep{miyake_etal16}.

One of the most remarkable effects is on the migration of low-mass planets (type I migration). \citet{ogihara_etal15a,ogihara_etal15b} performed \textit{N}-body simulations of protoplanets and/or planetesimals in disks evolving via disk winds, which is based on \citet{suzuki_etal10}, and found that type I migration is suppressed over the whole close-in region when the mass loss due to disk winds is relatively strong. They also demonstrated that type I migration can be even reversed in some cases. In addition, results of \textit{N}-body simulations indicate that disk winds would help in reproducing observed distributions of terrestrial planets including extrasolar super-Earths.

In recent disk evolution models (\citealt{bai_etal16}; \citealt{suzuki_etal16}), the effect of disk wind torque (magnetic braking) is also investigated. The disk winds carry away the angular momentum from the disk, which causes mass accretion. This is investigated in magneto-centrifugal disk wind model (e.g., \citealt{blandford_payne82}; \citealt{bai_stone13}; \citealt{simon_etal13}; \citealt{gressel_etal15}). 
\citet{suzuki_etal10} considered the mass loss due to disk winds alone, and ignored the effect of disk wind torque. \citet{suzuki_etal16} took this into account and calculated the global disk evolution. They found that the wind torque can alter the structure of gas surface density and flow, and in some cases, the radial mass accretion is dominated not by turbulence-driven accretion but by wind-driven accretion. Taking into account the wind-driven accretion, we can explain the observed mass accretion rate onto the central star. The observationally inferred mass accretion rate is larger than $10^{-9} M_\odot~{\rm yr}^{-1}$ at $t=10^6~{\rm yr}$, which was not explained by MRI-inactive disks. However, it is shown that the accretion rate via magnetic braking may reproduce the observed rate \citep{suzuki_etal16} even in MRI-inactive disks.

As an advance on our previous papers \citep{ogihara_etal15a,ogihara_etal15b}, here we incorporate two effects of disk winds. First, the evolution of gas surface density was calculated according to \citet{suzuki_etal16}, which considered the effect of wind torque. Second, effects of global gas flows on type I migration were considered. Global gas flows can be induced, for example, by wind-driven accretion, which may alter the picture of type I migration. According to recent studies of type I migration, the direction and rate of migration depend on saturation level of the corotation torque (e.g., \citealt{paardekooper_etal11}). When there exists an accretion flow, differences in radial velocity between planets and disk gas would change the contribution of corotation torque (e.g., \citealt{masset_papaloizou03}; \citealt{paardekooper14}). If there are high-velocity gas flows, the corotation torque may change. However, previous models did not examine these effects.

In this paper, we have investigated the effects of global gas flows on type I migration. In particular, we examined the case in which gas flow structures are developed by wind-driven accretion. We note that one can use the same recipe for more general gas flows. For example, the accretion flows would be vertically non-uniform in a protoplanetary disk with an MRI-dead zone, namely a layered accretion takes place (e.g., \citealt{gammie96}; \citealt{turner_sano08}; \citealt{flock_etal15}). In doing so, we kept the problem as simple as possible and develop a general analysis. Computationally expensive and challenging hydrodynamic simulations are not performed in this work. The plan of the paper is as follows. In Section~\ref{sec:wind_accretion} we describe the evolution of protoplanetary disks under the influence of wind-driven accretion; in Section~\ref{sec:gas_flow} we examine type I migration using dynamical torque model; in Section~\ref{sec:model_vis} we discuss effects of global gas flows on type I migration using a different approach (namely, a viscous diffusion approximation); in Section~\ref{sec:discussion} we discuss the outcomes of our simulations and give our conclusions.

\section{Effects of wind-driven accretion on mass accretion at midplane}
\label{sec:wind_accretion}

\begin{figure}
\resizebox{1.0 \hsize}{!}{\includegraphics{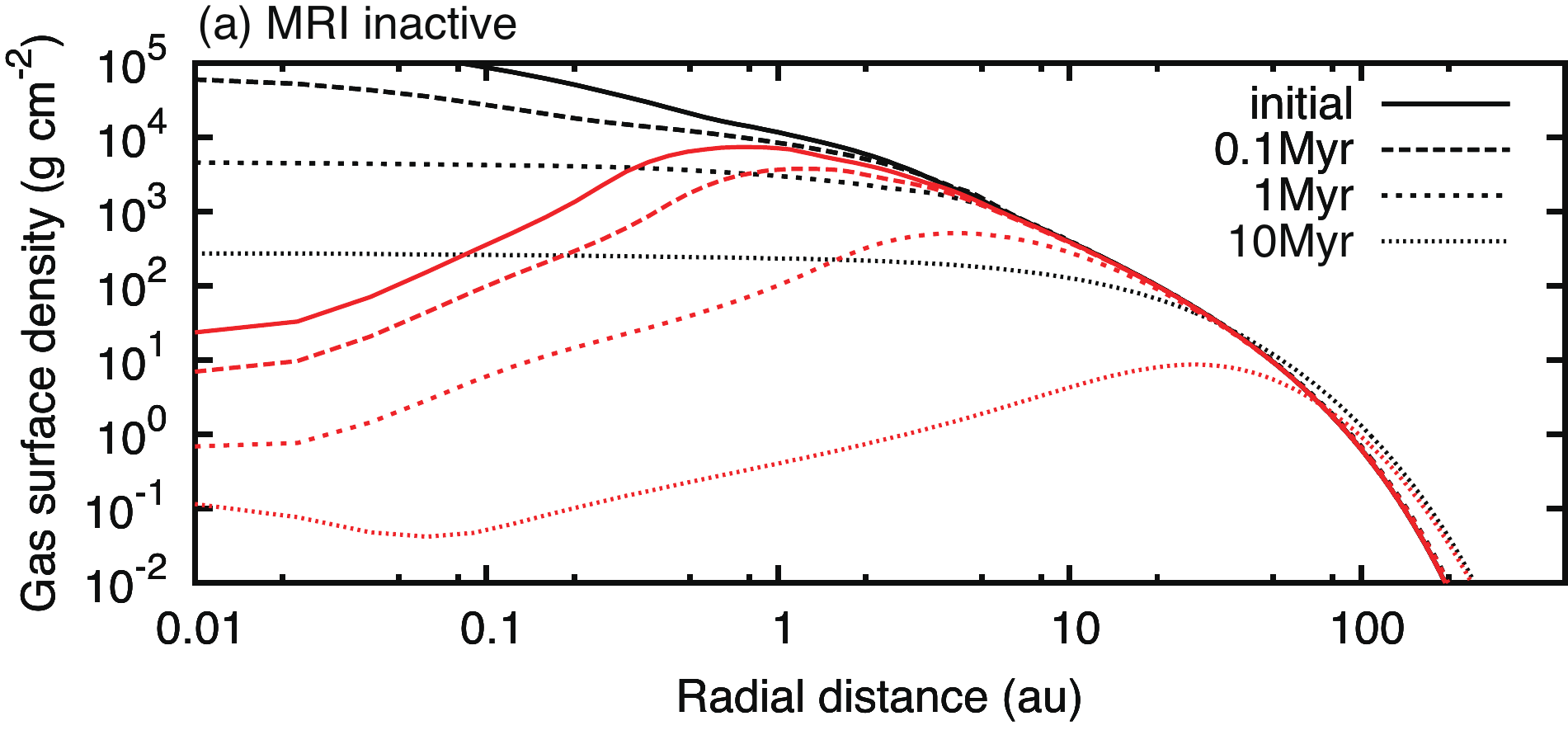}}
\resizebox{1.0 \hsize}{!}{\includegraphics{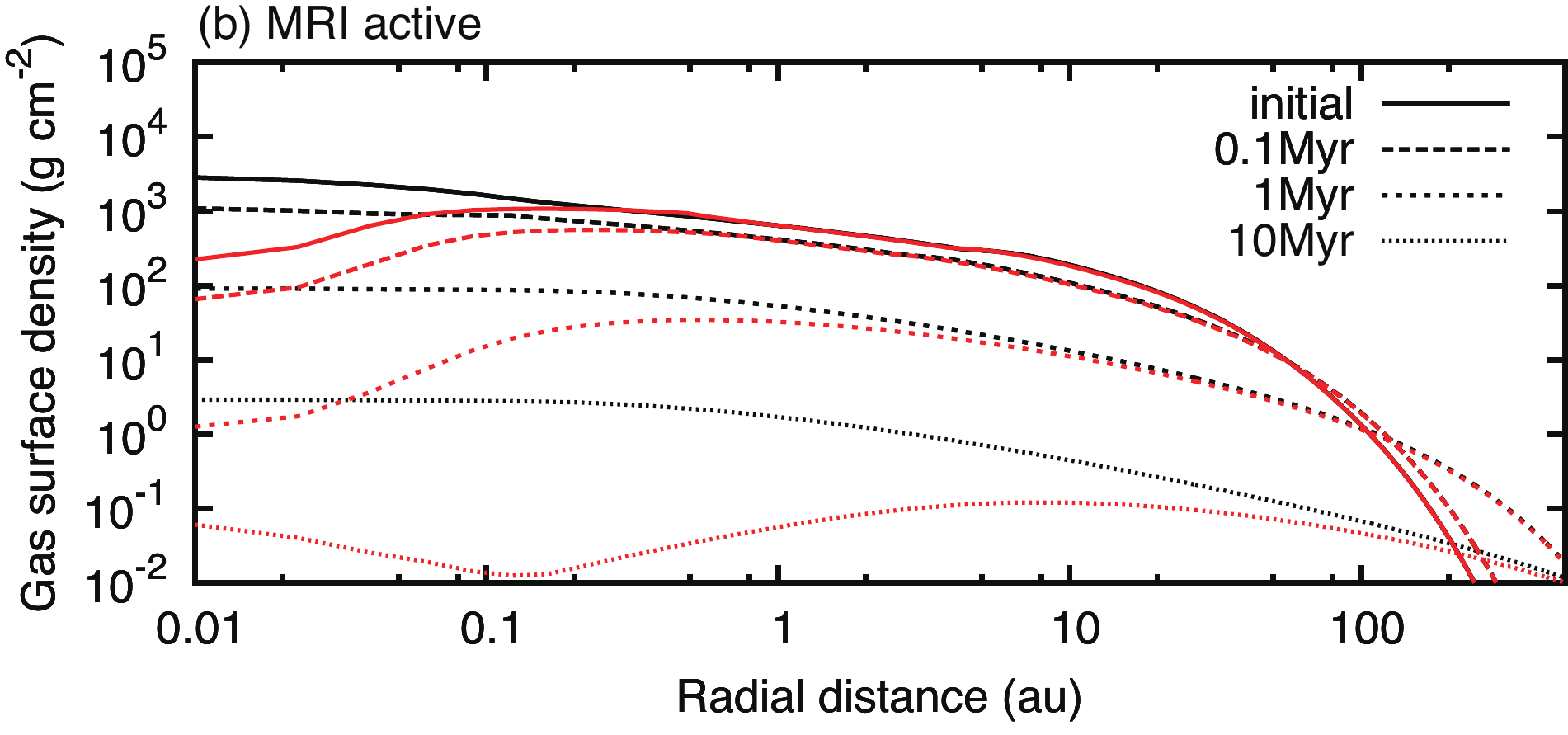}}
\caption{Time evolution of gas surface density of MRI-inactive disks ($\overline{\alpha_{r,\phi}} = 8.0 \times 10^{-5}$, panel (a)) and MRI-active disks ($\overline{\alpha_{r,\phi}} = 8.0 \times 10^{-3}$, panel (b)). The black lines represent disk evolution without wind-driven accretion (wind torque), while red lines indicate that with wind-driven accretion.
}
\label{fig:r_sigma}
\end{figure}

In our previous papers (\citealt{suzuki_etal10}; \citealt{ogihara_etal15a,ogihara_etal15b}), the evolution of protoplanetary disks is investigated with effects of mass loss due to disk winds without considering the wind torque (wind-driven accretion). In Figure~\ref{fig:r_sigma} we show the gas surface density evolution of MRI-inactive disks ($\overline{\alpha_{r,\phi}} = 8\times 10^{-5}$) and MRI-active disks ($\overline{\alpha_{r,\phi}} = 8\times 10^{-3}$) without considering wind torque, where $\overline{\alpha_{r,\phi}}$ indicates an effective turbulent viscosity, which is commonly referred to as the $\alpha$ parameter based on \citet{shakura_sunyaev73}. The disk evolution is the same as those shown in Figs.~1 and 5 in \citet{suzuki_etal16}. Their initial disk profile is set as a power-law distribution with a power-law index of -3/2 (Eq. 29 in \citealt{suzuki_etal16}), in which an exponential cutoff at $r = 30 {\rm ~au}$ is considered. We note that we are interested in orbital migration of planets in the type I regime and consider the stage when planets have grown to the isolation mass (e.g., \citealt{kokubo_ida98}). Taking into account the growth time of isolation-mass planets, we refer to $t=10^5$ years in \citet{suzuki_etal16} as initial $(t=0)$. We see that the surface density exhibits shallow negative slopes or even flat in inner region. We also found that type I migration can be suppressed in such region (e.g., \citealt{ogihara_etal15a,ogihara_etal15b}).

\citet{suzuki_etal16} developed a one-dimensional model of protoplanetary disks that evolve through magnetically driven disk winds. They also considered effects of the wind torque, in which wind carries away some angular momentum from the disk (also known as magnetic braking). This causes the radial mass accretion. Evolution of gas surface density including the effects of the wind torque is also shown in Fig.~\ref{fig:r_sigma}. We also refer the reader to Section~2 of \citet{suzuki_etal16} for details. The surface density slope is significantly different from the initial index of -3/2 in the close-in region ($r \lesssim 1 {\rm au}$) both for active and inactive disks. The slope can even be positive. Thus we can say that the wind torque increases the surface density slope. In MRI-inactive disks with smaller $\overline{\alpha_{r,\phi}}$, the positive surface density slope is more likely to be realized. In the region where the mass transport due to disk winds dominates over that due to viscous transport, the surface density slope tends to be positive values.

The surface density profile is altered only in the inner region, because the mass transport and mass loss due to disk winds are stronger in regions closer to the star. This is due to the fact that disk evolution time is proportional to the dynamical time at each location $r$, that is $r^{-3/2}$, and shorter for the inner region. 

\begin{figure}
\resizebox{1.0 \hsize}{!}{\includegraphics{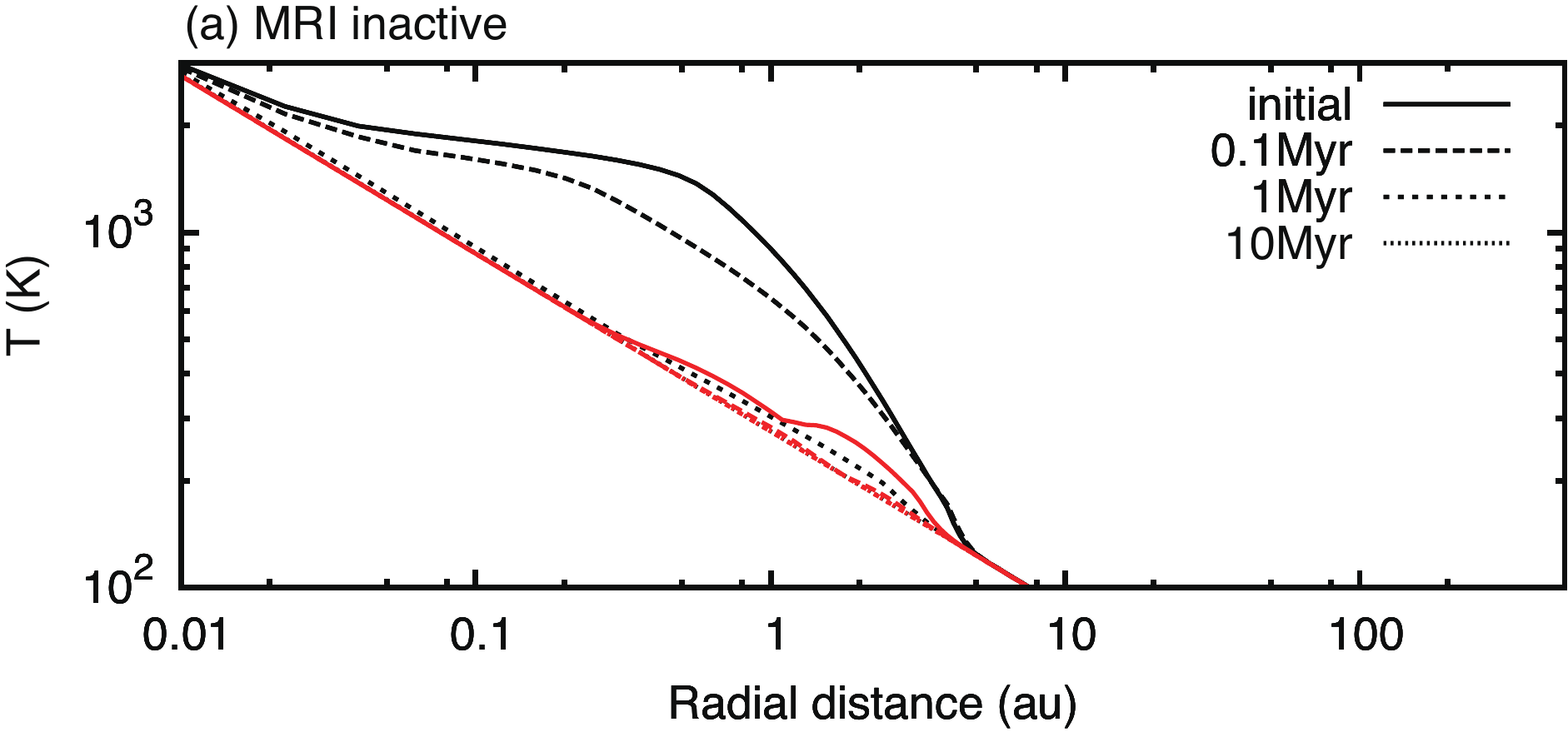}}
\resizebox{1.0 \hsize}{!}{\includegraphics{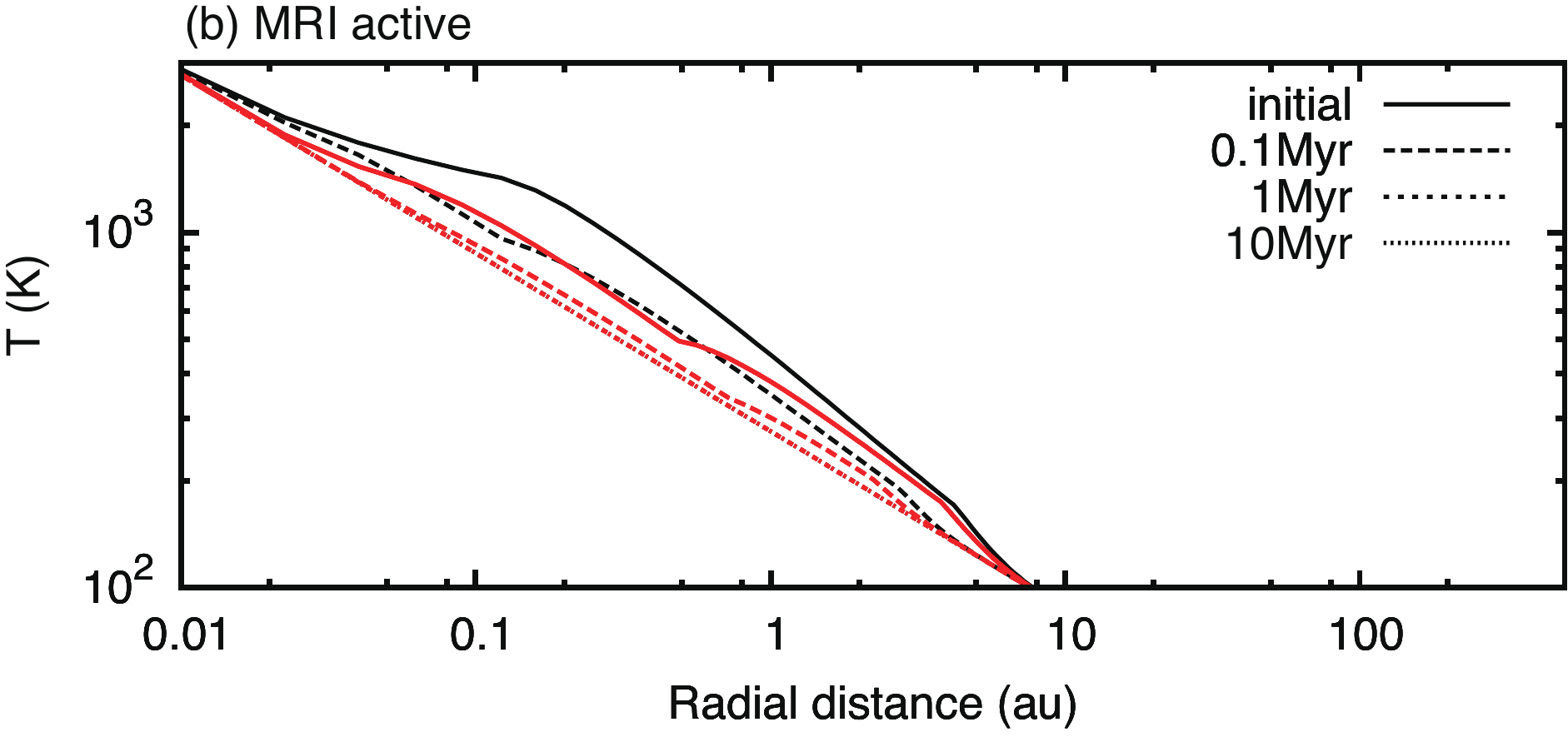}}
\caption{Time evolution of temperature of MRI-inactive disks ($\overline{\alpha_{r,\phi}} = 8.0 \times 10^{-5}$, panel (a)) and MRI-active disks ($\overline{\alpha_{r,\phi}} = 8.0 \times 10^{-3}$, panel (b)). The black lines represent disk evolution without wind-driven accretion (wind torque), while red lines indicate that with wind-driven accretion.
}
\label{fig:r_T}
\end{figure}

Figure~\ref{fig:r_T} shows the temperature evolution, which is the same as that in Figs.~1 and 5 of \citet{suzuki_etal16}. A detailed description of the temperature model is given in Section~2.4 of \citet{suzuki_etal16}, but here both viscous heating and stellar irradiation are considered. When the wind torque is not taken into account, the gas surface density is high in the inner region during the early phase of disk evolution. Thus, the disk temperature is determined by the viscous heating at $r < 10 {\rm ~au}$ and $t < 0.1 {\rm ~Myr}$, and the temperature slope is different from that of the radiative equilibrium for the minimum mass solar nebula \citep{hayashi81}. In the late phase of disk evolution without the wind torque and in almost all phases for a model with the wind torque, the temperature slope is considered to be $-1/2$.

\begin{figure}
\resizebox{1.0 \hsize}{!}{\includegraphics{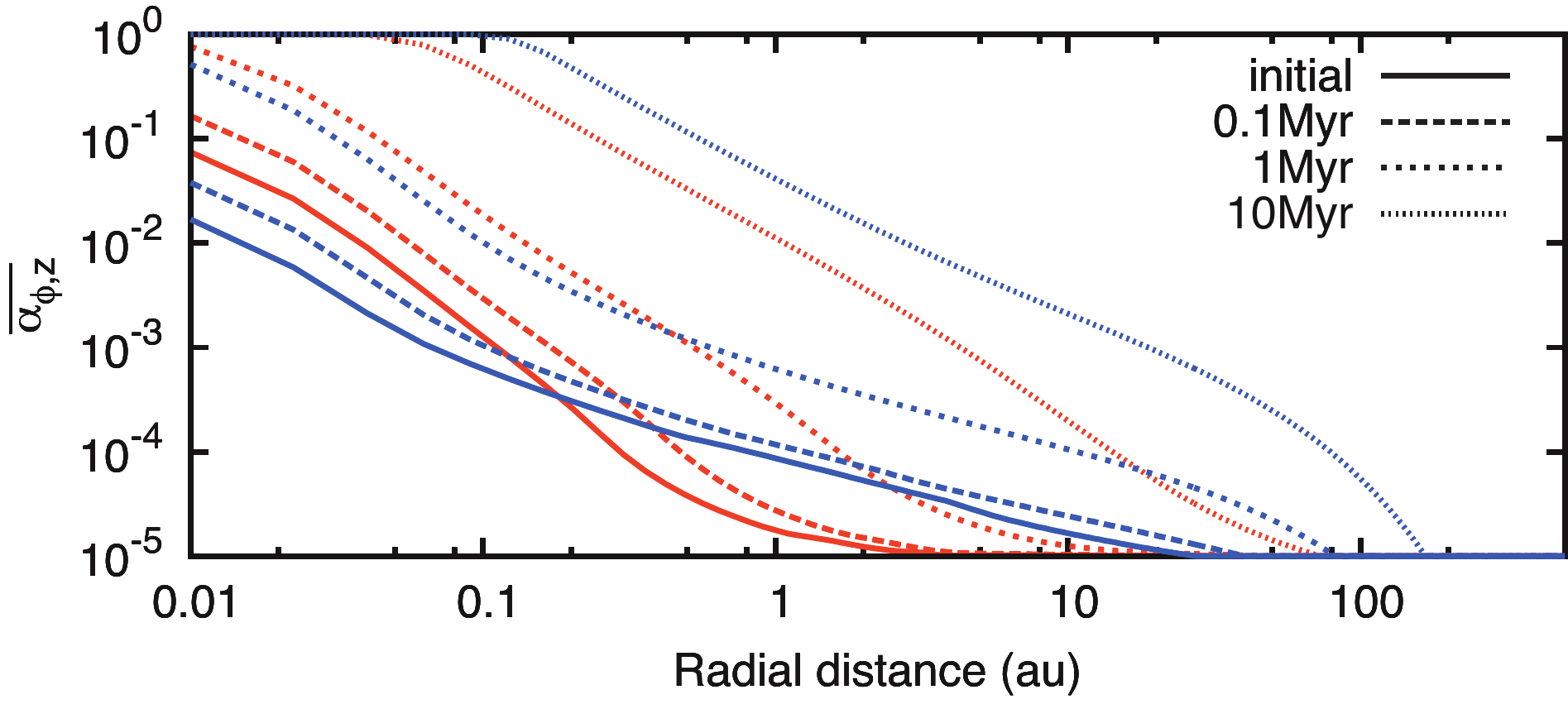}}
\caption{Time evolution of $\overline{\alpha_{\phi,z}}$ of MRI-inactive disks ($\overline{\alpha_{r,\phi}} = 8.0 \times 10^{-5}$, red) and MRI-active disks ($\overline{\alpha_{r,\phi}} = 8.0 \times 10^{-3}$, blue).
}
\label{fig:r_alpha}
\end{figure}

Figure~\ref{fig:r_alpha} indicates the evolution of $\overline{\alpha_{\phi,z}}$, which gives a measure of the angular momentum loss due to the wind torque. This is calculated according to \citet{bai13} and Eq.~(30) in \citet{suzuki_etal16}. That is,
\begin{eqnarray}
\label{eq:alpha_phi_z}
\overline{\alpha_{\phi,z}} = 10^{-5} \left(\frac{\Sigma_{\rm g}}{\Sigma_{\rm g,int}}\right)^{-0.66},
\end{eqnarray}
where $\Sigma_{\rm g,int}$ is the initial gas surface density given by Eq.~(29) in \citet{suzuki_etal16}. We see in Fig.~\ref{fig:r_alpha} that $\overline{\alpha_{\phi,z}}$ decreases with increasing $r$, which means that the global gas flow can be rapid in the inner disk\footnote{\citet{suzuki_etal16} put an upper limit on $\overline{\alpha_{\phi,z}} (<1)$. But, when $\overline{\alpha_{\phi,z}}\simeq 1$, accretion velocity $v_r$ can be close to the sound speed. Thus, it would be better to put smaller upper limit (i.e., $\overline{\alpha_{\phi,z}}\simeq0.1$). However, as shown in Fig.~\ref{fig:r_alpha}, $\overline{\alpha_{\phi,z}}$ is larger than 0.1 only in very close-in region (r<0.05 {\rm au}) after $t > 1 {\rm Myr}$. It is expected that type I migration is not affected by this assumption, because the gas density is already quite low there and planets in the region do not suffer type I migration.}. We used Eq.~(\ref{eq:alpha_phi_z}) up to $\overline{\alpha_{\phi,z}}\le 0.1$  by extending the original expression introduced by \citet{bai13} for $10^{-5}< \overline{\alpha_{\phi,z}}\le 10^{-3}$.
 
We cautiously note that the slopes of the surface densities that were obtained based on our simple model may be affected by various effects that are not considered here. First, we adopted the zero-torque-gradient condition, $\frac{\partial}{\partial r}(\Sigma_{\rm g} r^{3/2}) =0$, at the inner boundary, which forces the slope to be set to $\Sigma_{\rm g} \propto r^{-3/2}$ in the inner region. However, if a magnetospheric cavity is formed around the central star, the condition, $\Sigma_{\rm g} = 0$, would be more realistic for the inner boundary, as is adopted by, for example, \citet{matter_etal16}. In such circumstances, the positive slope of $\Sigma_{\rm g}$ would be enhanced in the inner region (Eq.~25 and Fig.~4 of \citealt{lynden-bell_pringle74}). 

Second, we assumed a spatially constant $\overline{\alpha_{r,\phi}}$, and that $\overline{\alpha_{\phi,z}}$ depends only on $\Sigma_{\rm g}$ (Eq.~\ref{eq:alpha_phi_z}). If the ionization effect by the radiation heating from the central star is included, both $\overline{\alpha_{r,\phi}}$ and $\overline{\alpha_{\phi,z}}$ are supposed to increase with $r$ for a ``usual'' negative slope of $\Sigma_{\rm g}$ because the radiation can penetrate toward the midplane for smaller $\Sigma_{\rm g}$, which is considered in the original expression of $\overline{\alpha_{\phi,z}}$ derived by \citet{bai13}. If the radial dependence of $\overline{\alpha_{r,\phi}}$, and $\overline{\alpha_{\phi,z}}$ is taken into account, the obtained positive slope of $\Sigma_{\rm g}$ would be suppressed at early time when the negative slope of $\Sigma_{\rm g}$ is still preserved. However, once the gas in the inner region is cleared out, we expect that the positive dependence of $\overline{\alpha_{r,\phi}}$ and $\overline{\alpha_{\phi,z}}$ on $r$ will be suppressed or even reversed.

\begin{figure}
\resizebox{1.0 \hsize}{!}{\includegraphics{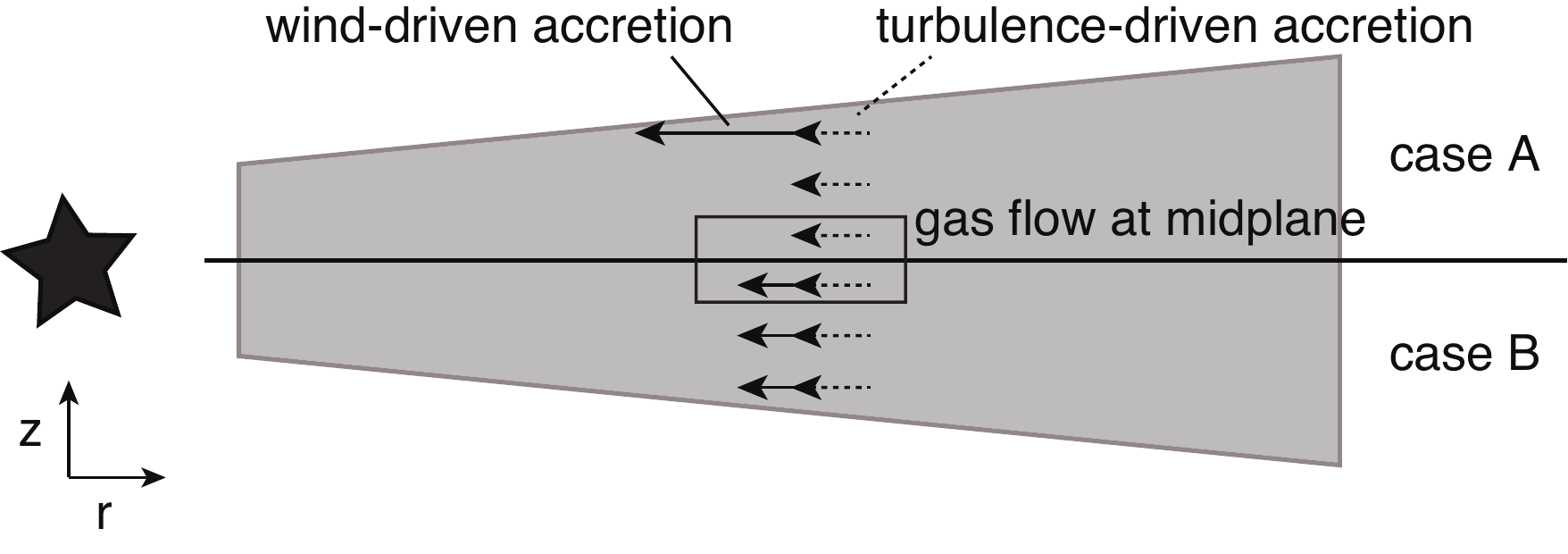}}
\caption{Schematic illustration of the vertical distribution of radial mass accretion for case~A and case~B. We note that the vertically integrated accretion rate is the same for each case. Thus, the surface radial velocity in case~A is generally much higher than throughout the disk's height in case~B.
}
\label{fig:wind_driven}
\end{figure}

There is an important caveat in the one-dimensional disk model; the vertical dependence of mass accretion is not known. Although the wind torque should operate at the disk surface, we do not know whether it affects the mass accretion at the midplane or not. We considered two cases. The first case is that the wind torque affects only the disk near the surface. At the midplane, the disk accretes solely through the turbulent viscosity ($v_r = v_{r,{\rm turb}}$). Hereafter we refer this situation as ``case~A.'' The second case is that the wind torque changes the accretion at the midplane and the mass accretion occurs over the entire region between midplane and surface ($v_r = v_{r,{\rm turb}} + v_{r,{\rm wind}}$). Hereafter we refer this situation as ``case~B.'' Case~A provides a rough lower limit on the mass accretion at the midplane, while case B corresponds to an upper limit. Figure~\ref{fig:wind_driven} shows a schematic picture of the vertical dependence of mass accretion for each case. The reality would lie between these two cases. To determine which cases represent the reality, detailed three-dimensional MHD simulations are required. Recently, three-dimensional global MHD simulations have been performed by various groups. Simulations with a net vertical magnetic flux show the accretion takes place near the surfaces, which can be modeled by our case~A, even without dead zones \citep{suzuki_inutsuka14, zhu_stone17}. On the other hand, simulations without a net vertical flux show the opposite radial flow pattern \citep{flock_etal11,fromang_etal11}. Vertical streams such as meridional circulation further perturb the radial flows and make the flow pattern more complicated \citep{bethune_etal17}. At the moment, the situation is far from conclusive. Instead of taking into account the effect of complicated flow structure, in this paper we consider each end-member case and examine planet migration in that framework.

The effect of global gas flows on type I migration have not been examined. Although we especially focused on the wind-driven accretion in this study, our model can be applied to more general flows. For example, case~A resembles a case with layered accretion. Again, we needed hydrodynamical simulations to assess this effect. As a first step, in this study, we adopted a simple recipe. We considered that streamlines are altered by the mass accretion onto the central star, which causes an imbalance of corotation torques, and investigate type I migration. We note that we considered type I migration of planets with small inclinations ($\ll 1$) and focus on motion near the midplane.

\section{Dynamical torque model}
\label{sec:gas_flow}

\subsection{Model}

We treat the gas drift as the feedback effect of the relative motion of the gas and planet. \citet{masset_papaloizou03} considered the relative motion between the gas and planet and found that a positive feedback exists as the corotation torque is proportional to the relative velocity, possibly leading to a runaway migration (type III migration). They suggested that the runaway migration happens for intermediate-mass planets. \citet{paardekooper14} proposed that a similar feedback effect can occur for low-mass planets, which they call dynamical corotation torque, and developed a formula of the torque. The type I migration torque is expressed by superposition of the static torque and dynamical corotation torque:
\begin{eqnarray}
\label{eq:gamma}
\Gamma = \Gamma_{\rm static} + \Gamma_{\rm dynamic},
\end{eqnarray}
where the static torque is the commonly used type I migration torque. For detailed description of the torque formulae, see Eqs.~(50)-(53) in \citet{paardekooper_etal11}. The dynamical torque is given by
\begin{eqnarray}
\label{eq:dynamic}
\Gamma_{\rm dynamic} = 2 \pi \left( 1- \frac{w_{\rm c}}{w(r)} \right) \Sigma_{\rm g} r^2 x_{\rm s} \Omega v_{\rm rel},
\end{eqnarray}
where $x_{\rm s}$ and $\Omega$ are the half-width of the horseshoe region and the Keplerian frequency, respectively, and $v_{\rm rel} = v_{r,{\rm planet}} - v_{r,{\rm gas}}$ is the relative velocity between the gas and planet. The quantity $w_{\rm c}$ indicates the inverse of the specific vorticity at the initial location of the planet, while $w(r)$ is the inverse of the specific vorticity at $r$. The specific vorticity is defined as the vertical component of the flow vorticity divided by the gas surface density: $(\nabla \times {\bm v})_z / \Sigma_{\rm g}$. As stated in \citet{paardekooper14}, the factor $1 - w_{\rm c}/w(r)$ plays the same role as the coorbital mass deficit in classical type III migration. This factor can be calculated by (\citealt{paardekooper14})
\begin{eqnarray}
1 - \frac{w_{\rm c}}{w(r)} = \left(\frac{3}{2} + p \right) \frac{x_{\rm s}^2}{6 r \nu}v_{\rm rel},
\end{eqnarray}
where $p = (d\ln \Sigma_{\rm g} / d\ln r)$ and $\nu$ represents the viscosity.

Using this approach, special care must be taken. First, \citet{paardekooper14} developed the dynamical torque formula assuming that the drift timescale across the horseshoe region is longer than the libration time ($x_{\rm s}/v_{\rm rel} > \tau_{\rm lib}$). The libration time is longer for smaller planets, so that this condition is not satisfied for such planets as will be seen in Section~\ref{sec:map2}. When this condition is violated, it is suggested that planets undergo runaway migration (\citealt{masset_papaloizou03}; \citealt{paardekooper14}; \citealt{pierens15}). However, it is uncertain whether such low-mass planets ($\lesssim 1 M_\oplus$) can undergo runaway migration. Instead, a cut-off of the corotation torque may be expected when the radial drift time across the horseshoe region is shorter than  the libration time \citep{masset01,masset02}. Thus we set $\Gamma_{\rm dynamic} \to 0$ for $x_{\rm s}/v_{\rm rel} < \tau_{\rm lib}$. Second, the dynamical torque formula (Eq.~(\ref{eq:dynamic})) is derived for locally isothermal disks. So we do not know the actual dynamical torque in radiative disks. In this study, the barotropic part of the corotation torque has a large value; thus the torque formula can be a good approximation. However, this would have to be taken into account in future work. 

The relative velocity between the gas and planet $v_{\rm rel}$, which was used in the calculation of the dynamical corotation torque, for cases~A and B are indicated below. 

\subsubsection{Case A: no wind-driven accretion at midplane}
\label{sec:caseA}

In case~A, the velocity of gas flow is determined by the turbulence-driven accretion:
\begin{equation}
\label{eq:v_turb}
v_{r,{\rm turb}} \simeq - \frac{1}{2}  \overline{\alpha_{r,\phi}} \left(\frac{H}{r}\right)^2 r \Omega,
\end{equation}
where $H$ is the disk scale hight.

\subsubsection{Case B: wind-driven accretion at midplane}
\label{sec:caseB}
In case~B, the radial gas velocity at the midplane is expressed by the sum of turbulence-driven accretion and wind-driven accretion ($v_{r,{\rm gas}} = v_{r,{\rm turb}} + v_{r,{\rm wind}}$). The velocity due to the turbulence-driven accretion is given by Eq.~(\ref{eq:v_turb}). The radial velocity due to the wind torque is (Eq.~34 in \citealt{suzuki_etal16}):
\begin{equation}
\label{eq:v_wind0}
v_{r,{\rm wind}} \simeq - \sqrt{\frac{2}{\pi}} \overline{\alpha_{\phi,z}} c_{\rm s},
\end{equation}
where $c_{\rm s}$ indicates the sound speed.

\subsection{Migration map}
\label{sec:map2}
\begin{figure}
\resizebox{1.0 \hsize}{!}{\includegraphics{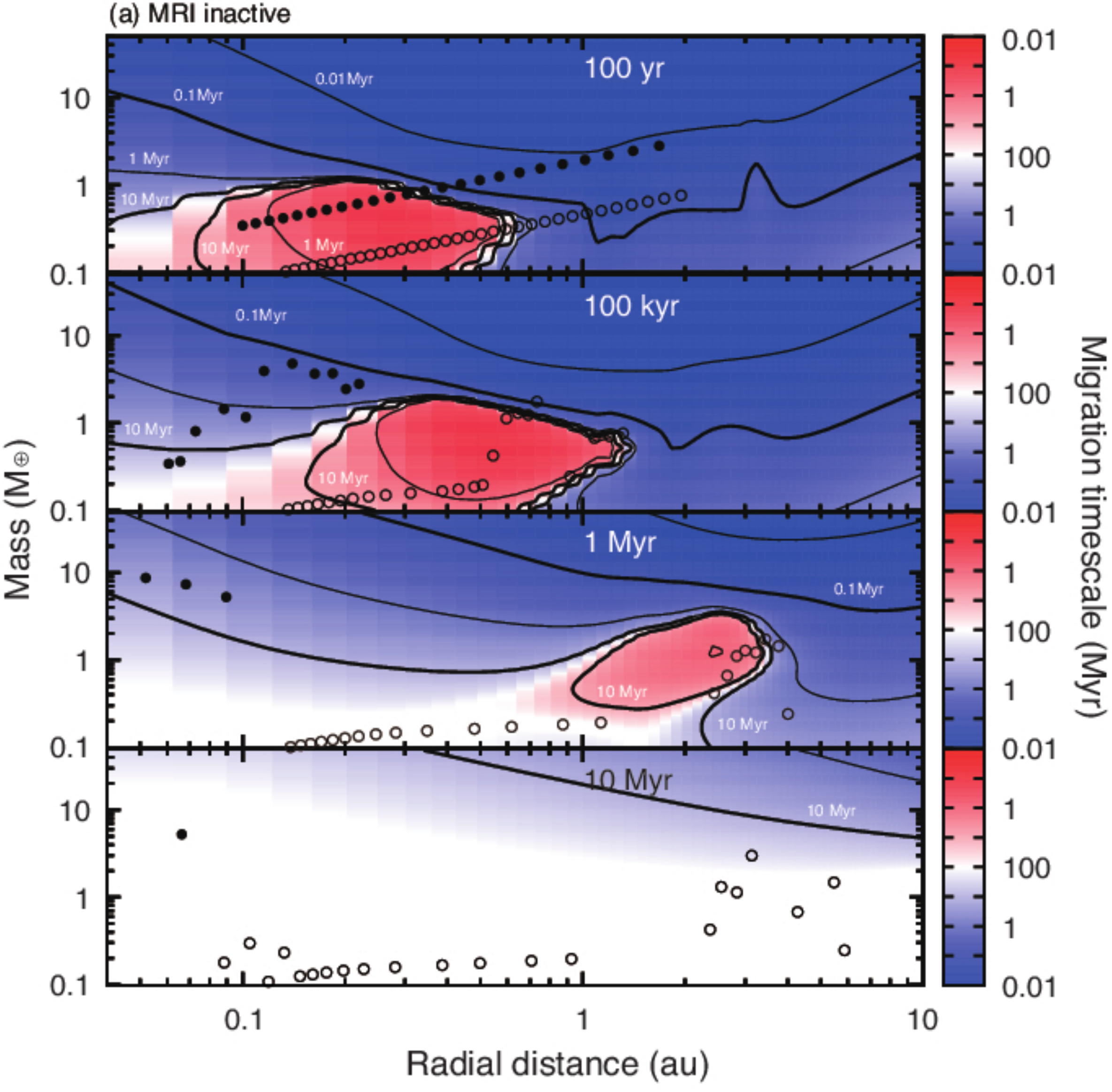}}
\resizebox{1.0 \hsize}{!}{\includegraphics{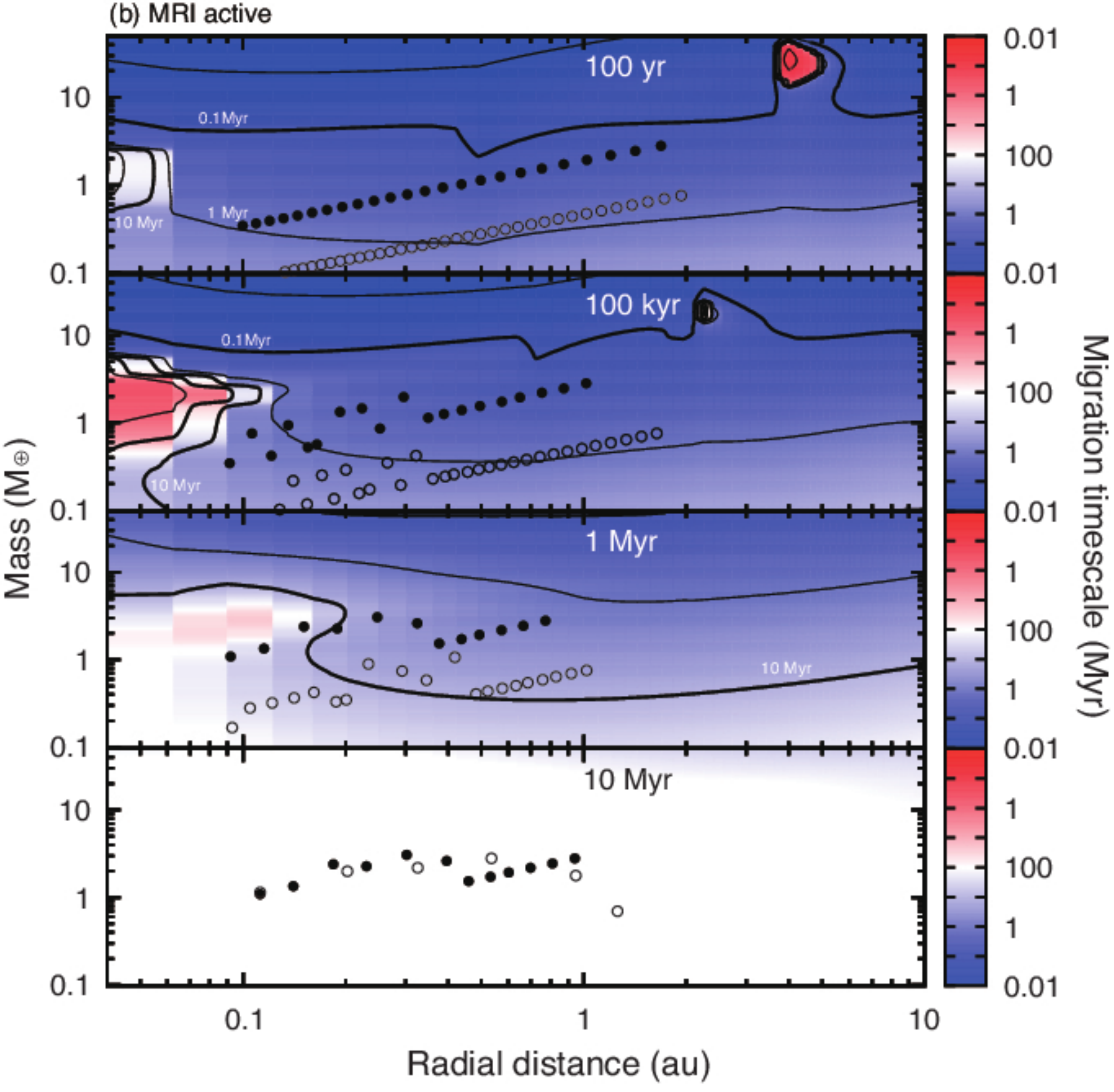}}
\caption{Migration maps of MRI-inactive disks ($\overline{\alpha_{r,\phi}} = 8.0 \times 10^{-5}$) and MRI-active disks ($\overline{\alpha_{r,\phi}} = 8.0 \times 10^{-3}$) based on case~A prescription. The color indicates the migration timescale. In the blue region, planet migration is inward, while in the red region planets undergo outward migration. Snapshots of systems of \textit{N}-body simulations are also shown by open circles ($\Sigma_0 = 30$) and filled circles ($\Sigma_0 = 75$), respectively. 
}
\label{fig:map17}
\end{figure}

\begin{figure}
\resizebox{1.0 \hsize}{!}{\includegraphics{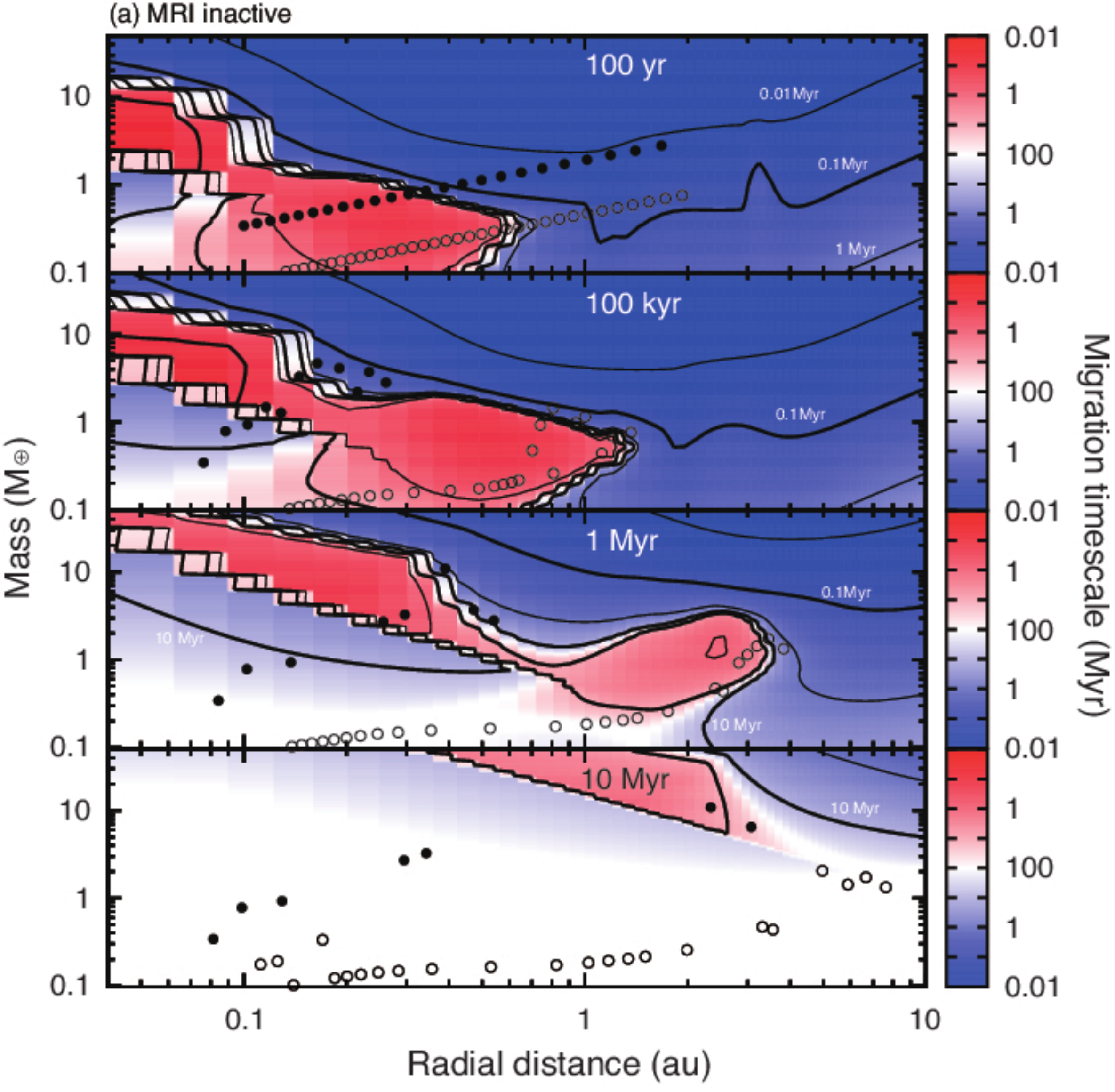}}
\resizebox{1.0 \hsize}{!}{\includegraphics{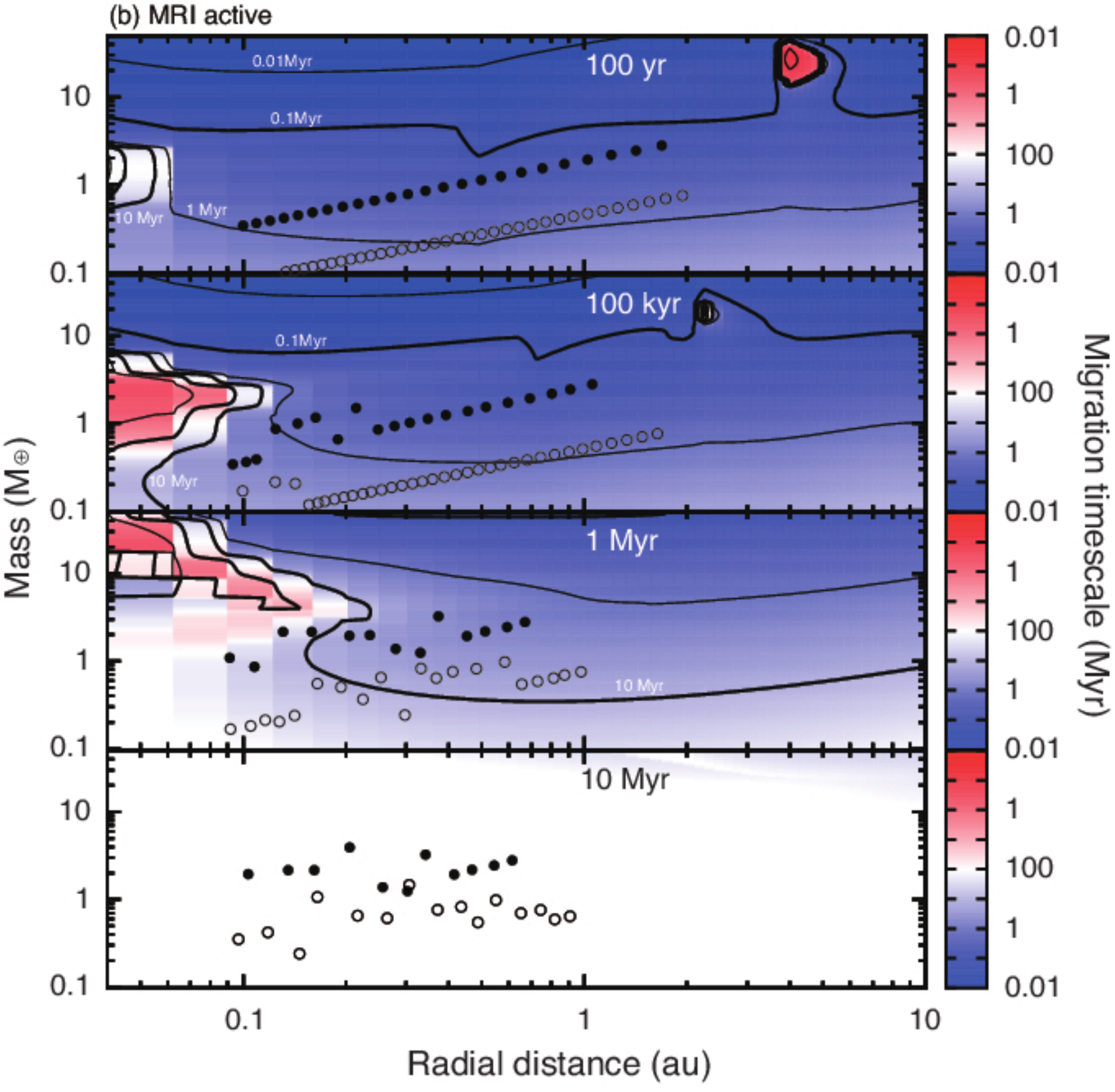}}
\caption{Same as Fig.~\ref{fig:map17} but based on case~B prescription. 
}
\label{fig:map21}
\end{figure}

Figure~\ref{fig:map17} shows the migration map when the midplane gas flow is solely determined by $\overline{\alpha_{r,\phi}}$ (case~A; Section~\ref{sec:caseA}), while Figure~\ref{fig:map21} indicates the map including the wind-driven accretion at the midplane (case~B; Section~\ref{sec:caseB}). In Figures~\ref{fig:map17}(a) and \ref{fig:map21}(a), $\overline{\alpha_{r,\phi}} = 8.0 \times 10^{-5}$ is assumed (MRI inactive), while in Figures~\ref{fig:map17}(b) and \ref{fig:map21}(b), $\overline{\alpha_{r,\phi}} = 8.0 \times 10^{-3}$ is used (MRI active). Here we assumed $v_{r,{\rm planet}} = 0$ when drawing the map. The color indicates the migration timescale with inward migration, outward migration, and no migration regions all shown. As described in Eq.~(\ref{eq:gamma}), the type I migration torque consists of the static torque and dynamical corotation torque. The static type I migration torques is expressed by the superposition of Lindblad and corotation torques. The Lindblad torque is basically negative, but the corotation torque can be positive depending on the vortensity gradient and the entropy gradient  of disk gas in the horseshoe region. As shown in Eqs.~(4)-(7) in \citet{paardekooper_etal11}, the barotropic part of corotation torque depends on the surface density slope, while the entropy-related corotation torque depends on the temperature slope and the surface density slope. In general, when the surface density slope is positive and the temperature slope is negative, the gas in the horseshoe region provides a positive torque, which can overwhelm the negative Lindblad torque.
Panels for $t = 100 {\rm ~yr}$ indicates the initial state.
Table~\ref{tbl:list} shows the summary of the model.

\begin{figure}
\resizebox{1.0 \hsize}{!}{\includegraphics{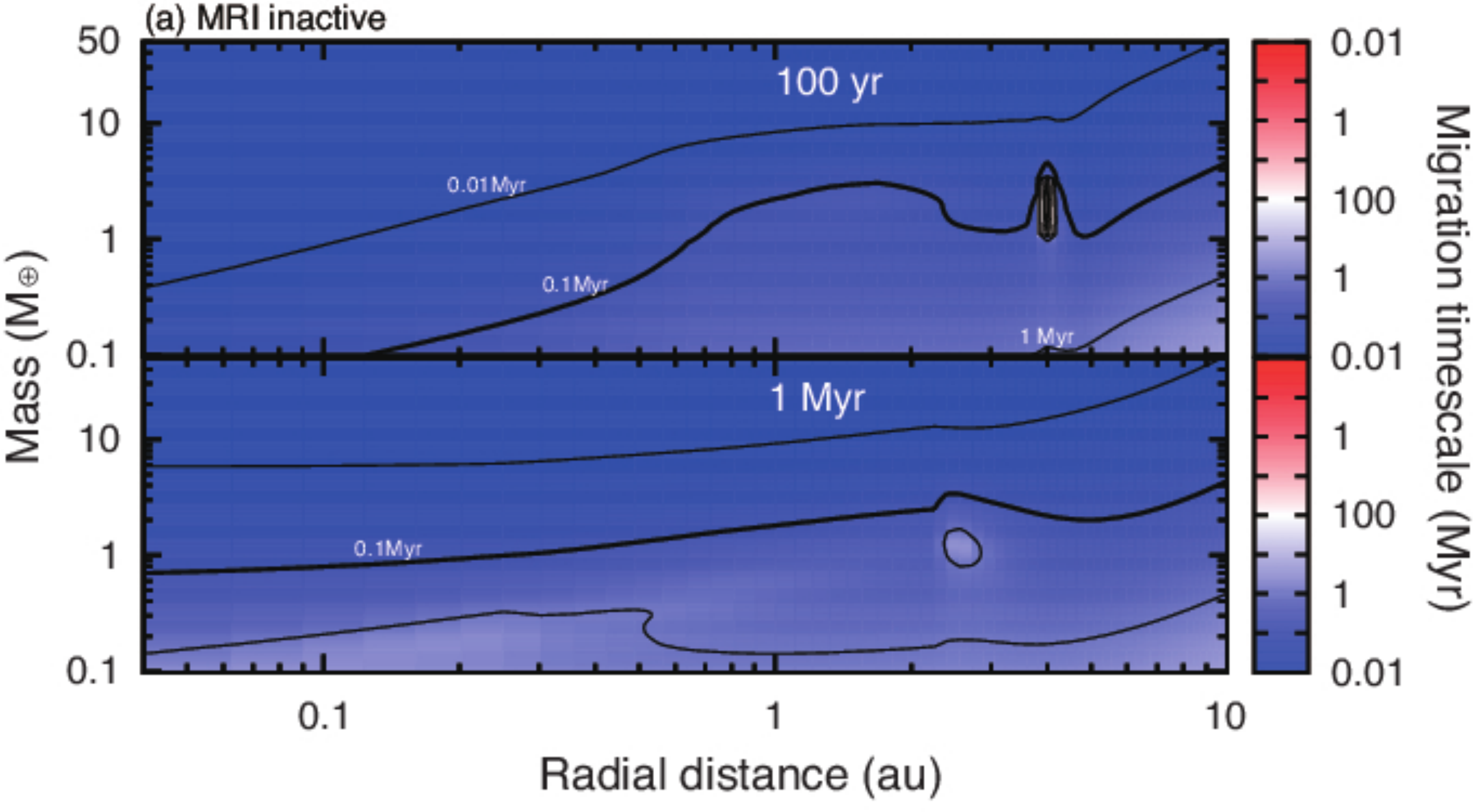}}
\resizebox{1.0 \hsize}{!}{\includegraphics{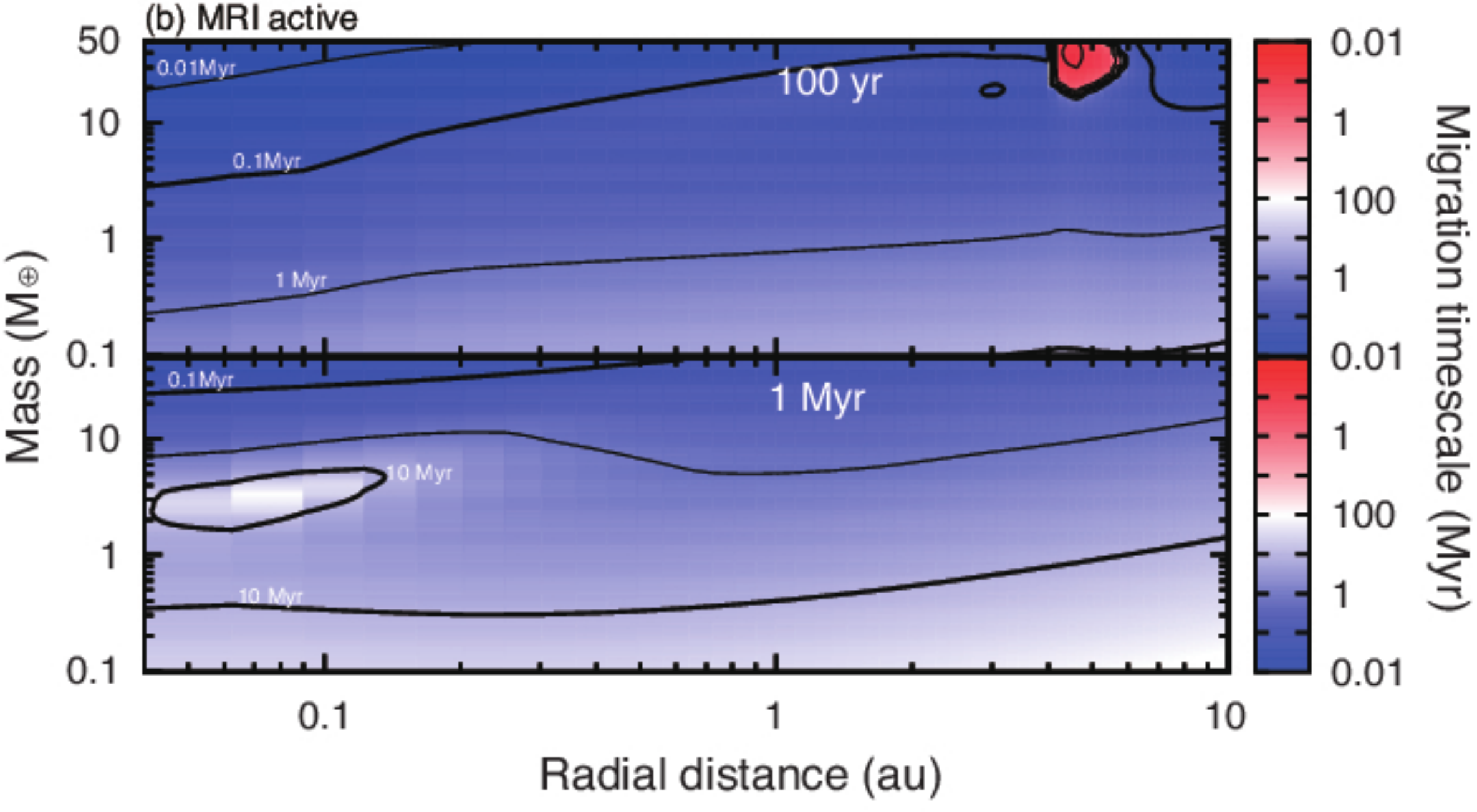}}
\caption{Migration map of MRI-inactive disks ($\overline{\alpha_{r,\phi}} = 8.0 \times 10^{-5}$) and MRI-active disks ($\overline{\alpha_{r,\phi}} = 8.0 \times 10^{-3}$), in which effects of wind torque are not considered for calculation of gas surface density eovlution. Corresponding disk evolution is shown in Fig.~\ref{fig:r_sigma} by black lines.
}
\label{fig:map427}
\end{figure}

For comparison, we plot the migration map for cases with mass loss due to disk winds but without wind-driven accretion due to the wind torque in Figure~\ref{fig:map427}. The migration map shows that inward migration is expected in almost the whole region. The inward migration timescale is longer than in the case with power-law density slope of -3/2 (see also \citealt{ogihara_etal15a,ogihara_etal15b}). In the inner region ($r \lesssim 1 {\rm ~au}$), the surface density profile is shallower than $-3/2$ and the barotropic part of the corotation torque partially compensate for the Lindblad torque. In addition, Figure~\ref{fig:r_T} shows that the temperature profile has a steep slope near $r \sim 1 {\rm au}$ for $t < 0.1 {\rm Myr}$ in the case without wind torque, especially for inactive disks. The entropy-related corotation torque has a positive value near $r \sim 1 {\rm au}$, which also partially compensates for the Lindblad torque. The total torque is still negative, leading to inward migration.
Comparing Figs.~\ref{fig:map17} and \ref{fig:map427}, effects of the wind torque on surface density slope (and hence on type I migration) are seen. When the wind torque is considered, the surface density is carved out from inside (see Fig.~\ref{fig:r_sigma}), so that the corotation torque compensates for the Lindblad torque and the disk dissipation timescale is shortened in the inner region, leading to a further suppression of inward migration. That is, the inward migration region in Fig.~\ref{fig:map427} changes to no migration or outward migration region in Fig.~\ref{fig:map17}. The temperature profile is altered for $t < 0.1 {\rm Myr}$ in the case without wind torque, which gives positive entropy-related corotation torque. In the case with wind torque, however, the temperature profile can be considered as a power-law distribution with index $-1/2$. Although the temperature profile is different, the main differences in migration map (see outward migration region in Fig.~\ref{fig:map17}) can be attributed to the difference in the surface density slope.

Comparing Figs.\ref{fig:map17}(a) and (b), we see that the region of outward migration is wider in the first case (MRI inactive disk). This is because, as seen in Figure~\ref{fig:r_sigma}, the region where the surface density slope is positive is wider in MRI-inactive disks than in MRI-active disks. As stated above, the positive surface density slope gives positive corotation torque. Thus the outward migration region is wider in Figures~\ref{fig:map17}(a) and \ref{fig:map21}(a) than in Figures~\ref{fig:map17}(b) and \ref{fig:map21}(b). As seen in Fig.~\ref{fig:r_T}, the temperature distribution can be considered as a power-law distribution, so that the temperature slope does not play a major role in changing the corotation torque. We note that in MRI active disks, the surface density slope is almost flat between $r = 0.1 - 1 $ au (see Fig.~\ref{fig:r_sigma}(b)), which results in substantial suppression of both inward and outward migration. We note also that in disks with no effects of disk winds, the outward migration region moves toward the lower-left region (e.g., \citealt{bitsch_etal15}). However, we see in migration maps that the outward migration region moves toward the right of the figure. This is because the region with a positive surface density slope expands to the outer region as time increases.
We also note that in case A, the direction and rate of type I migration are almost solely determined by the static torque, because the radial gas velocity at the midplane is not large enough to change the migration map.

Figure~\ref{fig:map21} shows the migration map, in which wind-driven accretion at the midplane is considered (case~B). That is, $v_{\rm rel} = \mathrm{d}a/\mathrm{d}t - (v_{r,{\rm turb}} + v_{r,{\rm wind}})$. We see that the outward migration region extends toward the upper-left region (smaller $r$ and larger $M$). 
As seen in Fig.~\ref{fig:r_alpha}, the radial velocity due to wind-driven accretion is large in inner region, leading to a high $v_{\rm rel}$. The magnitude of dynamical corotation torque is proportional to $v_{\rm rel}^2$, thus the torque is large in inner region. In the lower-left corner of the panels of Fig.~\ref{fig:map21}(a), the condition $x_{\rm s}/v_{\rm rel}>\tau_{\rm lib}$ is not satisfied and $\Gamma_{\rm dynamic}$ is set to zero.
Another important point concerns the sign of the dynamical torque. As discussed in \citet{paardekooper14}, the sign of the dynamical torque depends on the background vortensity gradient. If the surface density slope is positive, the dynamical torque has a positive value (and fully unsaturated corotation torque tends to be positive). 
We see no significant change in Figure~\ref{fig:map21}(b) from Figure~\ref{fig:map17}(b) before $t = 1 {\rm ~Myr}$. This is because the disk evolution is dominated by the turbulence-driven accretion in MRI-active disks ($\overline{\alpha_{r,\phi}} \simeq 10^{-2}$).

\begin{table}
\caption{Summary of models.}
\label{tbl:list}
\centering
\begin{tabular}{l l l l}
\hline\hline
case & $\overline{\alpha_{r,\phi}}$ &	migration map & orbital evolution\\
\hline 
case~A&	inactive& Fig.~\ref{fig:map17}(a)& Fig.~\ref{fig:t_a1}(a)\\
case~A&	active& Fig.~\ref{fig:map17}(b)& --\\
case~B&	inactive& Fig.~\ref{fig:map21}(a)& Fig.~\ref{fig:t_a1}(b)\\
case~B&	active& Fig.~\ref{fig:map21}(b)& --\\
\hline
\end{tabular}
\end{table}

\begin{figure}
\resizebox{0.9 \hsize}{!}{\includegraphics{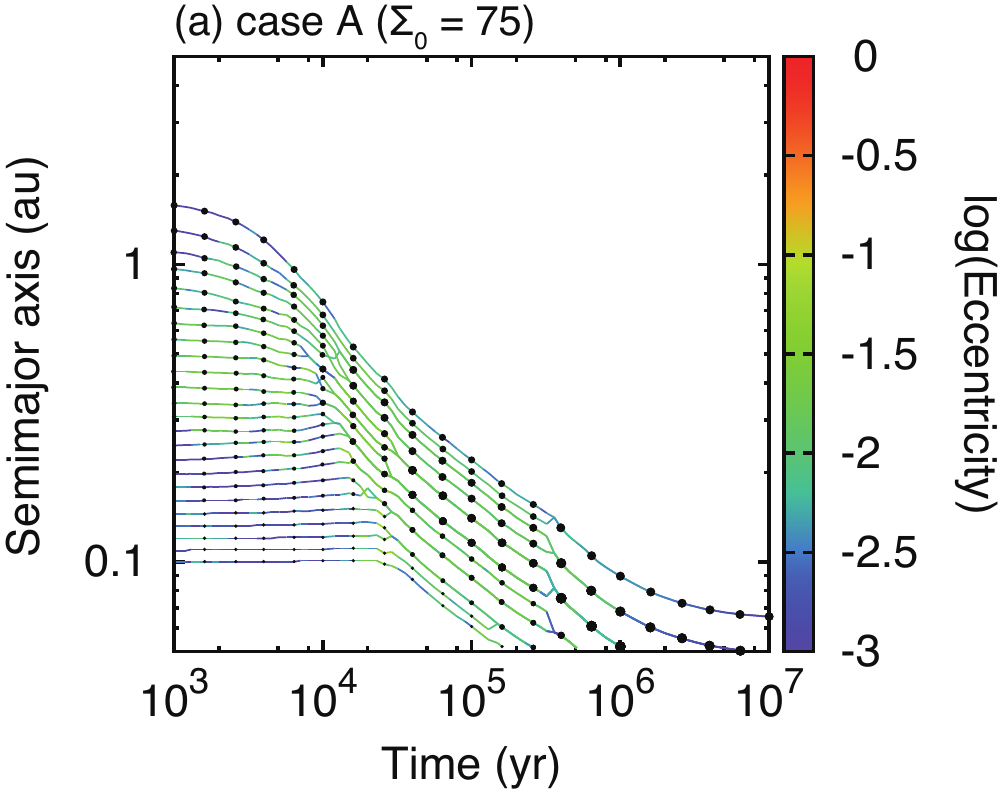}}
\resizebox{0.9 \hsize}{!}{\includegraphics{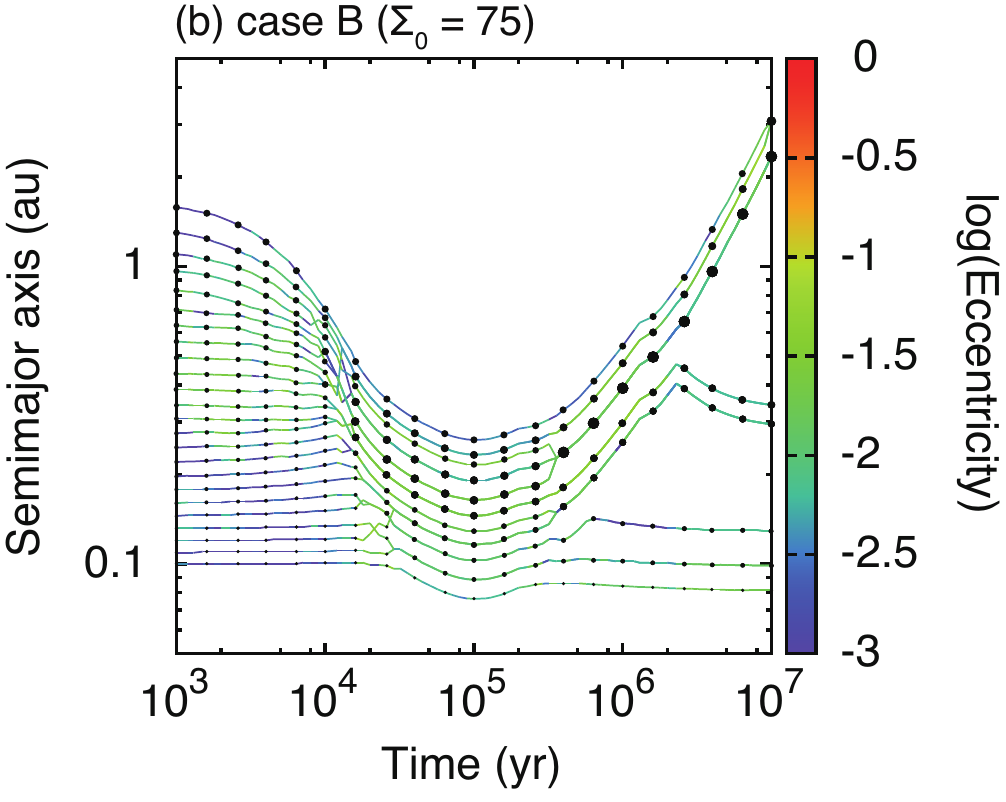}}
\caption{Orbital evolution of planets for $\Sigma_0 = 75$ in MRI-inactive disks. Snapshots of the system for panels (a) and (b) are shown by filled circles in Fig.~\ref{fig:map17}(a) and Fig.~\ref{fig:map21}(a), respectively. The filled circles connected with solid lines represent the sizes of planets. The color of the lines indicates the eccentricity (color bar).
}
\label{fig:t_a1}
\end{figure}

\subsection{Orbital evolution}
In order to demonstrate the effects on planet formation, we perform \textit{N}-body simulations of orbital evolution of isolation-mass planets. Planets undergo the type I migration and the tidal damping of eccentricities and inclinations (see Eq.~6 of \citealt{ogihara_etal15a} for $e$-damping timescale). We note that, although $v_{r,{\rm planet}}=0$ is used when drawing migration maps in Figs.~\ref{fig:map17} and \ref{fig:map21}, the actual radial velocity of the planet is used in calculating the dynamical torque. For numerical integration, we use a fourth-order Hermite scheme with a hierarchical individual time step. When the physical radii of two spherical bodies overlap, they are assumed to merge into one body, assuming perfect accretion. The initial planetary mass $(M_{\rm iso} = 2\pi a \Delta a \Sigma_{\rm d})$ is calculated according to power-law solid distributions:
\begin{eqnarray}
\Sigma_{\rm d} = \Sigma_0 \left(\frac{r}{1{\rm au}}\right)^{-3/2} {\rm g~cm^{-2}},
\end{eqnarray}
where $\Sigma_0 = 30$ or 75 is used. Then
\begin{eqnarray}
M_{\rm iso} \simeq 0.15 \left(\frac{\Delta a}{10~r_{\rm H}} \right)^{3/2} \left(\frac{a}{1~{\rm au}} \right)^{3/4} \left(\frac{\Sigma_0}{10~{\rm g~cm^{-2}}} \right)^{3/2}~M_\oplus.
\end{eqnarray}
The initial separation between oligarchs is set by $\Delta a = 10 r_{\rm H}$, where $r_{\rm H}$ is the mutual Hill radius. Snapshots of each system are overplotted in Figs.~\ref{fig:map17} and \ref{fig:map21}. Open circles represent the case in which $\Sigma_0 = 30$, while solid circles represent the case for $\Sigma_0 = 75$.

As seen in Figs.~\ref{fig:map17}(b) and \ref{fig:map21}(b), results are similar between cases~A and B in MRI-active disks because migration maps are similar. The surface density slope is almost flat in the inner region (Fig.~\ref{fig:r_sigma}(b)) and migration is significantly suppressed in MRI-active case.

From Figures~\ref{fig:map17}(a) and \ref{fig:map21}(a), we see that orbital evolution and final orbital configurations differ in each model. In case~B, no planets migrate inside $r = 0.08$ au (Fig.~\ref{fig:map21}(a)). In case~A, planets of a few Earth masses pass over the outward migration zone (red region in Fig.~\ref{fig:map17}(a)). This does not occur in case~B, because the outward migration zone extends to high-mass regions (red region in Fig.~\ref{fig:map21}(a)), which prevents the inward migration of super-Earth mass planets. Thus wind-driven accretion has an effect of preventing inward migration. In addition, in Fig.~\ref{fig:map21}(a), super-Earth mass planets undergo outward migration to outer region ($r>5 {\rm au}$) (filled circles). For more detailed time evolution of semimajor axis for $\Sigma_0 = 75$ in MRI-inactive disks, see Fig.~\ref{fig:t_a1}(a) (case~A) and (b) (case~B). As stated above, no inward migration is observed in Fig.~\ref{fig:t_a1}(b).

According to \citet{paardekooper14}, planets may undergo runaway migration when the $k$-coefficient defined by Eq.~(31) of \citet{paardekooper14} is larger than 0.5. This condition for runaway migration is satisfied only for $r < 0.1 {\rm ~au}$ and $M \gtrsim 5 ~M_\oplus$ at $t = 0.1 {\rm ~Myr}$ in Fig.~\ref{fig:map21}(a). In our \textit{N}-body simulations, no planets satisfy this condition.

\section{Discussion and conclusions}

\subsection{A different approach --diffusion model}
\label{sec:model_vis}

Before we make concluding remarks, we briefly see effects of global gas flows on migration based on a different approach. According to \citet{paardekooper_etal11}, the corotation torque for type I migration is expressed by using the following parameters:
\begin{eqnarray}
P_\nu = \frac{2}{3} \left( \frac{\Omega r^2 \tilde{x}_{\rm s}^3}{2 \pi \nu} \right)^{1/2},\\
P_\chi = \left( \frac{\Omega r^2 \tilde{x}_{\rm s}^3}{2 \pi \chi} \right)^2 = \frac{3}{2} P_\nu \left( \frac{\nu}{\chi} \right)^{1/2},
\end{eqnarray}
where $\tilde{x}_{\rm s}$, $\nu$ and $\chi$ are dimensionless half-width of the horseshoe region, viscous and thermal coefficients, respectively. 
Values of effective viscosity $\nu$ is presented below. For thermal diffusion coefficient $\chi$, we use the same form as Eq.~(34) of \citet{paardekooper_etal11}, in which a coefficient is corrected as described in \citet{bitsch_kley11}. The corotation torque has a maximum for $P_\nu \simeq 0.3$ (see Fig.~3 in \citealt{paardekooper_etal11}). A simple physical interpretation for the dependence on diffusivities is as follows. 
The corotation torque comes about because of the gas in the planet's horseshoe region making inward and outward U-turns relative to the planet at different vortensities and specific entropies. Both these quantities evolve in time because of their advection-diffusion-creation within the horseshoe region, and the magnitude of the corotation torque thus depends on how the timescales for viscous and thermal diffusions across the horseshoe region compare with the horseshoe U-turn and libration timescales \citep{baruteau_etal14}.

We calculated an effective ``viscosity'' from radial accretion and examine the consequent de-saturation of corotation torque. To incorporate effects of gas flows due to wind-driven accretion in the formula of type I migration, we derived an ``effective'' viscosity from the radial gas flow.

\subsubsection{Case A: no wind-driven accretion at midplane}

 In case~A, in which the viscosity at the midplane is solely expressed by $\overline{\alpha_{r,\phi}}$, the diffusion at the midplane is solely due to MRI turbulence ($\nu = \nu_{\rm turb}$):
\begin{equation}
\label{eq:nu1}
\nu_{\rm turb} \simeq \frac{2}{3} \overline{\alpha_{r,\phi}} \frac{c_{\rm s}^2}{\Omega}
= \frac{1}{3} \overline{\alpha_{r,\phi}} \left(\frac{H}{r}\right)^2 r^2 \Omega.
\end{equation}
Then we calculated the corotation torque by using
\begin{eqnarray}
P_\nu = \frac{2}{3} \left( \frac{\Omega r^2 \tilde{x}_{\rm s}^3}{2 \pi \nu_{\rm turb}} \right)^{1/2}.
\end{eqnarray}
In this case, we confirmed that the migration map is almost identical to that in Fig.~\ref{fig:map17} both for MRI-inactive and active disks.

\subsubsection{Case B: wind-driven accretion at midplane}

In case~B, using an approximate relation between $\nu$ and $v_r$ of
\begin{equation}
\label{eq:v_wind}
v_{r,{\rm wind}} \simeq - \frac{3}{2} \frac{\nu_{\rm wind}}{r},
\end{equation}
the effective ``viscous'' coefficient due to wind-driven accretion can be expressed as
\begin{equation}
\label{eq:nu2}
\nu_{\rm wind} \simeq \frac{2 \sqrt{2}}{ 3 \sqrt{\pi}} \overline{\alpha_{\phi,z}} r c_{\rm s} 
= \frac{2}{3 \sqrt{\pi}} \overline{\alpha_{\phi,z}} \left(\frac{H}{r}\right) r^2 \Omega,
\end{equation}
where Eq.~(\ref{eq:v_wind0}) is used for $v_{r,{\rm wind}}$.
Then we calculated the corotation torque using
\begin{eqnarray}
P_\nu = \frac{2}{3} \left[ \frac{\Omega r^2 \tilde{x}_{\rm s}^3}{2 \pi (\nu_{\rm turb} + \nu_{\rm wind})} \right]^{1/2}.
\end{eqnarray}

\begin{figure}
\resizebox{1.0 \hsize}{!}{\includegraphics{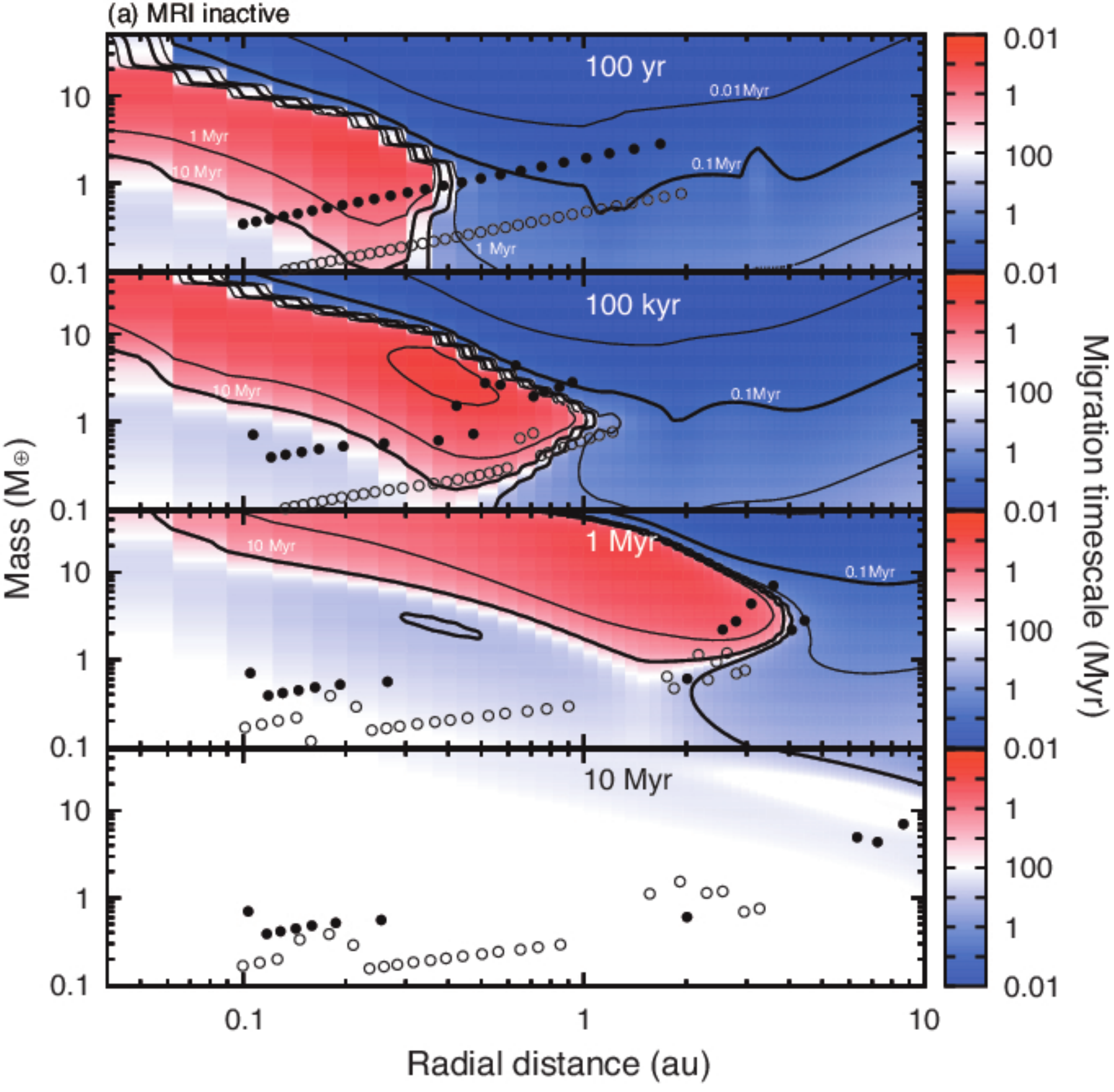}}
\resizebox{1.0 \hsize}{!}{\includegraphics{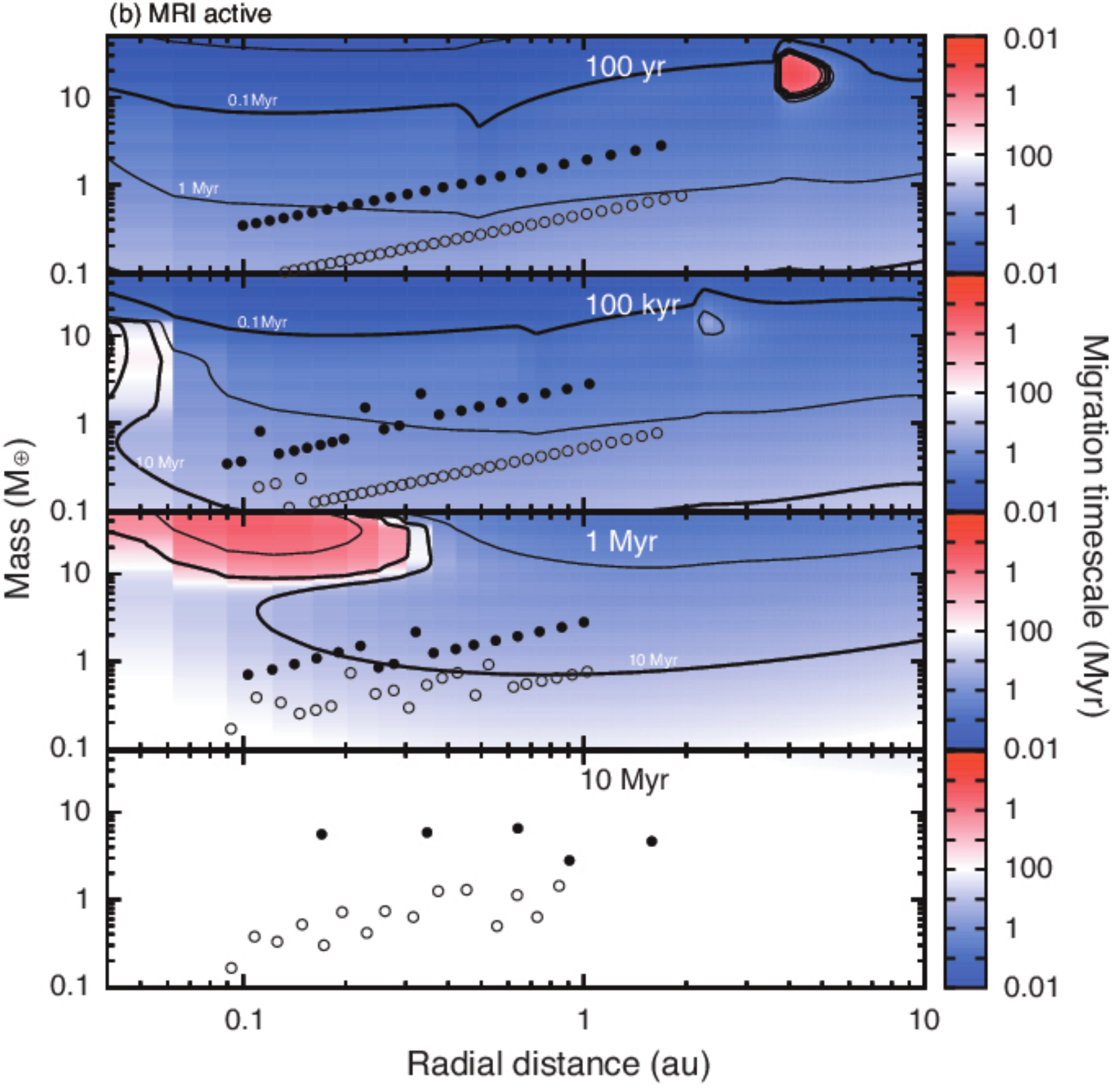}}
\caption{Same as Fig.~\ref{fig:map21} but based on a diffusion approach. 
}
\label{fig:map344}
\end{figure}

Figure~\ref{fig:map344} shows the migration map for case~B.
Regarding the effect of wind-driven accretion on desaturation of corotation torque, the outward migration region is more extended towards the upper-left region in Figure~\ref{fig:map344}(a). This is because the effective viscosity is larger in closer region due to high $\overline{\alpha_{\phi,z}}$. The corotation torque has a maximum for $P_\nu \simeq 0.3$, and this corresponds to planets with $M \simeq 6 ~M_\oplus$ ($t = 100 {\rm ~yr}$) and $M \simeq 30 ~M_\oplus$ ($t = 1{\rm ~Myr}$) at $r = 0.1 {\rm ~au}$ in Fig.~\ref{fig:map344}(a). However, for MRI-active disks, we see no significant differences between Figures~\ref{fig:map17}(b) and \ref{fig:map344}(b) before $t = 1$ Myr. This is because in MRI-active disks, $\overline{\alpha_{r,\phi}}$ is large and $\nu$ is determined primary by $\overline{\alpha_{r,\phi}}$ except for the very late stage ($t > 1 {\rm Myr}$). 

We see that the migration map of Fig.~\ref{fig:map21}(a) and \ref{fig:map344}(a) show similar trends.
The similarity might be explained by the fact that in the region of the disk with a positive surface density gradient the de-saturated corotation torque is positive \citep{masset_etal06}; likewise, the dynamical torque is positive, because the gas flows from a high vortensity region to a low vortensity region \citep{paardekooper14}.

One concern may be whether the gas flow promoted by the wind torque can be interpreted as a flow of the angular momentum into the corotation region, as presented in Section~\ref{sec:model_vis}. We think hydrodynamical simulations are required to determine if this approach is valid or not. However, the rapid gas flow at the surface due to wind-driven accretion induces a large velocity difference between the surface and the midplane, which would cause some diffusion processes (e.g., turbulence). In this case, our viscosity approach may represent reality. We note that it is not clear how likely it is that this turbulence can be interpreted as a source of diffusion, because this depends on the statistical properties of turbulence (e.g., \citealt{baruteau_lin10}). This should be investigated in future studies.

\subsection{Conclusions}
\label{sec:discussion}

We have examined effects of wind-driven accretion on type I migration. As seen in \citet{suzuki_etal16}, the disk wind torque has an effect of changing the surface density slope. The surface density slope can be carved out from inside.
In addition, we considered effects of gas flows on type I migration and found that gas flows at the midplane has an effect of preventing the infall of super-Earths onto the central star. In MRI-inactive disks, the surface density slope can be positive due to the wind torque and an outward migration region appears in the migration map. The outward migration region extends for case~B, in which the gas flow at the midplane is induced due to wind-driven accretion. This prevents the inward migration of massive super-Earths in the close-in region. We simulated orbital evolution of planets by \textit{N}-body simulations and observed that no planets fall onto the star in this case. We also see super-Earth mass planets migrate outward. In the MRI-active disks, we usually obtained a disk with flat slope in inner region due to the effect of wind torque, which results in substantial suppression of both inward and outward migration. In MRI-active case, the effect of gas flows due to wind-driven accretion $(\overline{\alpha_{\phi,z}})$ on desaturation of the corotation torque is not important, because the turbulence-driven accretion $(\overline{\alpha_{r,\phi}})$ primarily controls the surface density slope and corotation torque.

We note that although similar trends are observed in migration maps using different approaches, they are not perfectly matched with each other. One may also wonder that if planets open a density gap, the formulae of type I migration would not be valid and planets would migrate in the type II regime\footnote{Note that the strong corotation torque can also slow down type II migration in a disk with positive density slope \citep{crida_morbidelli07}.}. So, we checked the gap-opening criterion of \citet{crida_etal06} and found that even massive super-Earths do not open a gap. This is due to high mass accretion rate due to wind-driven accretion.

We note that the torque formula of \citet{paardekooper_etal11} would be accurate only when $(M/M_*)/(H/r)^3 < 1$. For example, planets with $M > 5 ~M_\oplus$ at $r = 0.1 {\rm ~au}$ and with $M > 40 ~M_\oplus$ at $r = 1 {\rm ~au}$ do not satisfy this condition. Although no planets in our \textit{N}-body simulations break this condition, the torque formula should be treated with some caution for more massive planets.

In this paper, we focused on wind-driven accretion; however one can apply the model to more general gas flows. For example, case~A of MRI-inactive disks would correspond to a dead zone region with an active surface layer.

As another application of this study, it is interesting to investigate how planets with high inclination undergo migration. The relative velocity between gas and planet changes with $z$, which may change migration. 
Special care would have to be taken in this case because the standard formula for type I migration is derived for $i < H/r$.

As \citet{suzuki_etal16} cautioned, the disk evolution with effects of magnetically-driven disk winds depends on the evolution of the net vertical magnetic field. It is very important to reveal the evolution of the vertical magnetic field by MHD simulations. From another perspective, effects of the magnetic filed on the corotation torque have also been investigated (e.g., \citealt{guilet_etal13}; \citealt{uribe_etal15}), which also depends on the evolution of the magnetic field.

It is very important to address two issues by performing hydrodynamical simulations in future work. First, we need to determine the vertical dependence of mass accretion when the wind-driven accretion is considered. Second, we need to examine how planets undergo type I migration when the radial mass accretion is considered. We would encourage (M)HD simulations that explore these issues.

\begin{acknowledgements}
We thank the anonymous referee for valuable comments. We also thank Bertram Bitsch, Hidekazu Tanaka and Takanori Sasaki for useful comments and suggestions. Numerical computations were conducted on the general-purpose PC farm at the Center for Computational Astrophysics, CfCA, of the National Astronomical Observatory of Japan. This work was supported by JSPS KAKENHI Grant Number 16H07415.
\end{acknowledgements}


{}


\begin{thebibliography}{}

\bibitem[Bai(2013)]{bai13}
Bai, X.-N. 2013,
\apj, 772, 96
\bibitem[Bai et al.(2016)]{bai_etal16}
Bai, X.-N. 2016,
\apj, 821, 80
\bibitem[Bai \& Stone(2013)]{bai_stone13}
Bai, X.-N., \& Stone, J. M. 2013,
\apj, 769, 76
\bibitem[Baruteau \& Lin(2010)]{baruteau_lin10}
Baruteau, C., \& Lin, D. N. C. 2010,
\apj, 709, 759
\bibitem[Baruteau et al.(2014)]{baruteau_etal14}
Baruteau, C., Crida, A., Paardekooper, S.-J., et al. 2014, 
in Protostars Planets VI, ed. H. Beuther et al. (Tucson, AZ: Univ. Arizona Press), 667
\bibitem[B\'ethune et al.(2017)]{bethune_etal17}
B\'ethune, W., Lesur, G., \& Ferreira, J. 2017,
\aap, 600, A75
\bibitem[Bitsch et al.(2015)]{bitsch_etal15}
Bitsch, B., Johansen, A., Lambrechts, M., \& Morbidelli, A. 2015,
\aap, 575, A28
\bibitem[Bitsch \& Kley(2011)]{bitsch_kley11}
Bitsch, B., \& Kley, W. 2011,
\aap, 536, A77
\bibitem[Blandford \& Payne(1982)]{blandford_payne82}
Blandford, R. D., \& Payne, D. G. 1982,
\mnras, 199, 883
\bibitem[Crida \& Morbidelli(2007)]{crida_morbidelli07}
Crida, A., \& Morbidelli, A. 2007,
\mnras, 377, 1324
\bibitem[Crida et al.(2006)]{crida_etal06}
Crida, A., Morbidelli, A., \& Masset, F. 2006,
\icarus, 181, 587
\bibitem[Dzyukevich et al.(2013)]{dzyukevich_etal13}
Dzyurkevich, N., Turner, N. J., Henning, Th., \& Kley, H. 2013, \apj, 765, 114
\bibitem[Flock et al.(2011)]{flock_etal11}
Flock, M., Dzyurkevich, N., Klahr, H., Turner, N. J., \& Henning, Th. 2011, \apj, 735, 122
\bibitem[Flock et al.(2012)]{flock_etal12}
Flock, M., Dzyurkevich, N., Klahr, H., Turner, N. J., \& Henning, Th. 2012, \apj, 744, 144
\bibitem[Flock et al.(2015)]{flock_etal15}
Flock, M., Ruge, J. P., Dzyurkevich, N., Henning, Th., Klahr, H., \& Wolf, S. 2015,
\aap, 574, A68
\bibitem[Fromang et al.(2011)]{fromang_etal11}
Fromang, S., Lyra, W., \& Masset, F. 2011, \aap, 534, A107
\bibitem[Gammie(1996)]{gammie96}
Gammie, C. F. 1996,
\apj, 457, 355
\bibitem[Gressel et al.(2015)]{gressel_etal15}
Gressel, O., Turner, N. J.; Nelson, R, P., \& McNally, C. P. 2015, \apj, 801, 84
\bibitem[Guilet et al.(2013)]{guilet_etal13}
Guilet, J., Baruteau, C., \& Papaloizou, J. C. B. 2013,
\mnras, 430, 1764
\bibitem[Kokubo \& Ida(1998)]{kokubo_ida98}
Kokubo E., \& Ida S. 1998,
\icarus, 131, 171
\bibitem[Lynden-Bell \& Pringle(1974)]{lynden-bell_pringle74}
Lynden-Bell, D., \& Pringle, J. E. 1974,
\mnras, 168, 603
\bibitem[Hayashi(1981)]{hayashi81}
Hayashi, C. 1981,
Prog. Theor. Phys. Suppl., 70, 35
\bibitem[Masset(2001)]{masset01}
Masset, F. S. 2001,
\apj, 558, 453
\bibitem[Masset(2002)]{masset02}
Masset, F. S. 2002,
\aap, 387, 605
\bibitem[Masset \& Papaloizou(2003)]{masset_papaloizou03}
Masset, F. S., \& Papaloizou, J. C. B. 2003,
\apj, 588, 494
\bibitem[Masset et al.(2006)]{masset_etal06}
Masset, F. S., Morbidelli, A., Crida, A., \& Ferreira, J. 2006,
\apj, 642, 478
\bibitem[Matter et al.(2016)]{matter_etal16}
Matter, A., Labadie, L., Augereau, J. C., Kluska, J., Crida, A., Carmona, A., Gonzalez, J. F., Thi, W. F., Le Bouquin, J. -B., Olofsson, J., \& Lopez, B. 2016,
\aap, 586, A11
\bibitem[Miyake et al.(2016)]{miyake_etal16}
Miyake, T., Suzuki, T. K., \& Inutsuka, S. 2016,
\apj, 821, 3
\bibitem[Ogihara et al.(2015a)]{ogihara_etal15a}
Ogihara, M., Kobayashi, H., Inutsuka, S., \& Suzuki, T. K. 2015,
\aap, 579, A65
\bibitem[Ogihara et al.(2015b)]{ogihara_etal15b}
Ogihara, M., Morbidelli, A., \& Guillot, T. 2015,
\aap, 584, L1
\bibitem[Paardekooper(2014)]{paardekooper14}
Paardekooper, S. -J. 2014,
\mnras, 444, 2031
\bibitem[Paardekooper et al.(2011)]{paardekooper_etal11}
Paardekooper, S. -J., Baruteau, C., \& Kley, W. 2011,
\mnras, 410, 293
\bibitem[Parkin(2014)]{parkin14}
Parkin, E. R. 2014, \mnras, 438, 2513
\bibitem[Parkin \& Biknell(2013)]{parkin_biknell13}
Parkin, E. R., \& Biknell, G. V. 2013, \apj, 763, 114
\bibitem[Pierens(2015)]{pierens15}
Pierrens, A. 2015,
\mnras, 454, 2003
\bibitem[Shakura \& Sunyaev(1973)]{shakura_sunyaev73}
Shakura, N. I., \& Sunyaev, R. A. 1973,
\aap, 24, 337
\bibitem[Simon et al.(2013)]{simon_etal13}
Simon, J. B., Bai, X.-N., Armitage, P. J., Stone, J. M., \& Beckwith, K. 2013, \apj, 775, 73
\bibitem[Suzuki \& Inutsuka(2009)]{suzuki_inutsuka09}
Suzuki, T. K., \& Inutsuka, S. 2009,
\apj, 691, L49
\bibitem[Suzuki et al.(2010)]{suzuki_etal10}
Suzuki, T. K., Muto, T., \& Inutsuka, S. 2010,
\apj, 718, 1289
\bibitem[Suzuki \& Inutsuka(2014)]{suzuki_inutsuka14}
Suzuki, T. K., \& Inutsuka, S. 2014, \apj, 784, 121
\bibitem[Suzuki et al.(2016)]{suzuki_etal16}
Suzuki, T. K., Ogihara, M., Morbidelli, A., Crida, A. \& Guillot, T. 2016,
\aap, 596, A74
\bibitem[Tanaka et al.(2002)]{tanaka_etal02} 
Tanaka, H., Takeuchi, T., \& Ward, W. R. 2002,
\apj, 565, 1257
\bibitem[Turner \& Sano(2008)]{turner_sano08}
Turner, N. J., \& Sano, T. 2008, \apjl, 679, L131
\bibitem[Uribe et al.(2015)]{uribe_etal15}
Uribe, A. L., Bans, A., \& K\"onigl, A. 2015,
\apj, 802, 54
\bibitem[Zhu \& Stone(2017)]{zhu_stone17}
Zhu, Z. \& Stone, J. M. 2017, arxiv:1701.04627
\end{thebibliography}

\end{document}